\documentclass[review,authoryear]{elsarticle}

\usepackage{geometry}
\usepackage[english]{babel}
\usepackage{amsmath}
\usepackage{amsthm}
\usepackage{amssymb}
\usepackage{colortbl}
\usepackage{chngcntr}
\usepackage{here}
\usepackage[toc,page]{appendix}
\usepackage{lineno}

\graphicspath{{figures/}}

\title{When very slow is too fast - collapse of a predator-prey system}

\author[1]{Anna Vanselow\corref{cor1}}
\ead{anna.vanselow@uni-oldenburg.de}
\author[2]{Sebastian Wieczorek}
\ead{sebastian.wieczorek@ucc.ie}
\author[1]{Ulrike Feudel}
\ead{ulrike.feudel@uni-oldenburg.de}

\cortext[cor1]{Corresponding author}
\address[1]{ICBM, Carl von Ossietzky University Oldenburg, Oldenburg, Lower Saxony, Germany}
\address[2]{University College Cork, Department of Applied Mathematics, Cork, Ireland}

\begin{document}

\begin{frontmatter}

\begin{abstract}
Critical transitions or regime shifts are sudden and unexpected changes in the state of an ecosystem, that are usually associated with {\em dangerous levels} of environmental change. However, recent studies show that critical transitions can also be triggered by {\em dangerous rates} of environmental change. 
In contrast to classical regime shifts, such {\em rate-induced critical transitions} do not involve any obvious loss of stability, or a bifurcation, and thus cannot be explained by the linear stability analysis. 
In this work, we demonstrate that the well-known Rosenzweig-MacArthur predator-prey model can undergo a rate-induced critical transition in response to a continuous decline in the habitat quality, resulting in a collapse of the predator and prey populations. Rather surprisingly,  the collapse occurs even if  the environmental change is slower than the slowest process in the model. To explain this counterintuitive phenomenon, we combine methods from {\em geometric singular perturbation theory} with the concept of a {\em moving equilibrium}, and study {\em critical rates} of environmental change with dependence on the initial state and the system parameters. Moreover, for a fixed rate of environmental change, we determine the set of initial states that undergo a rate-induced population collapse.
Our results suggest that ecosystems may be more sensitive to how fast environmental conditions change than previously assumed. In particular, unexpected critical transitions with dramatic ecological consequences can be triggered by environmental changes that (i) do not exceed any dangerous levels, and (ii) are slower than the natural timescales of the ecosystem. This poses an interesting research question whether regime shifts observed in the natural world are predominantly rate-induced  or bifurcation-induced.
\end{abstract}



\end{frontmatter}

\linenumbers
\modulolinenumbers[2]
\section{Introduction}

\begin{figure}[H]
\centering
\includegraphics[scale=0.6]{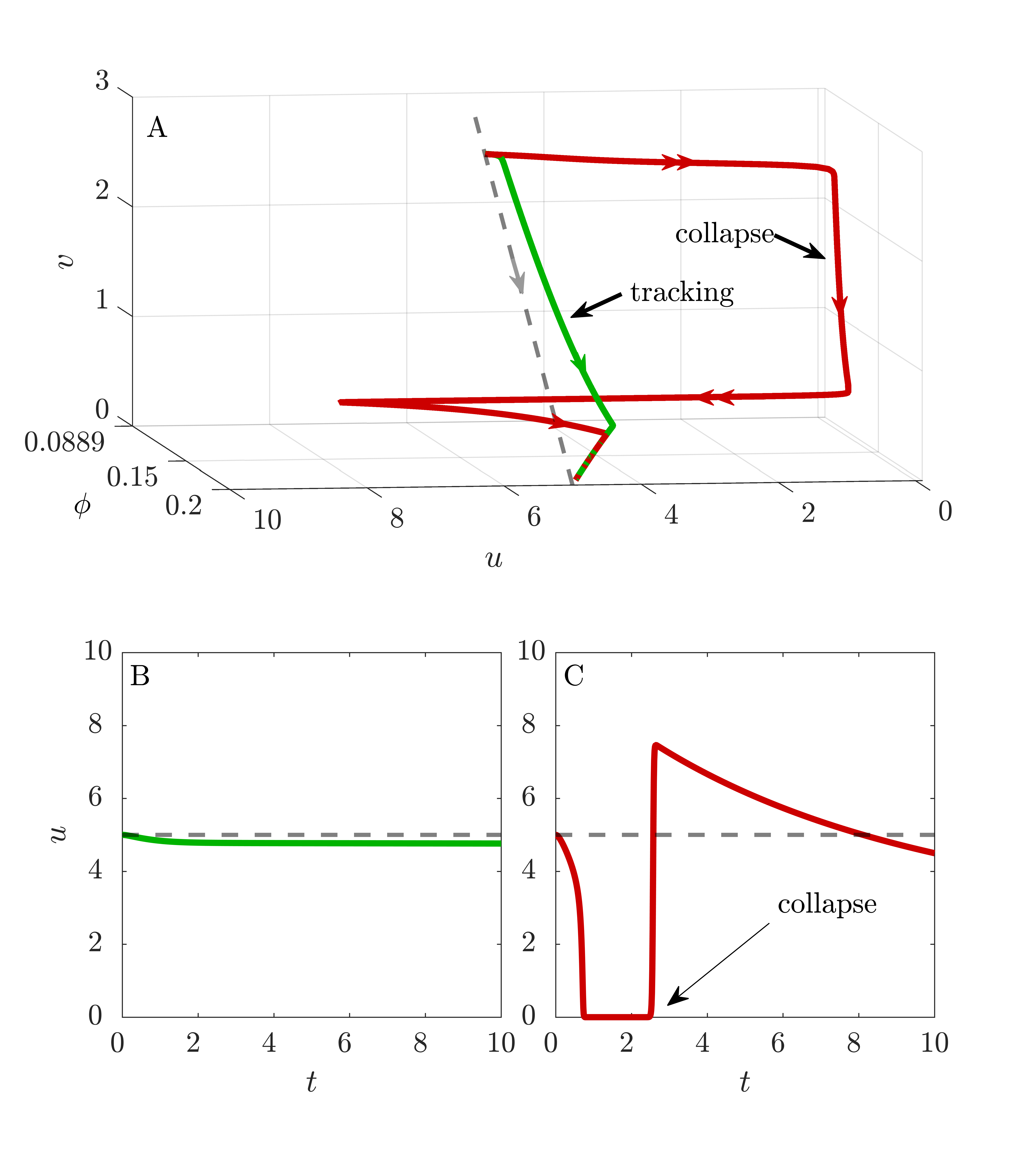}
\caption{(A): Three-dimensional phase portrait of the time-scaled Rosenzweig-MacArthur predator-prey system, with fast evolving prey $u$, slower reproducing predator $v$ and even slower changing environmental conditions $\phi$. Two different responses (red and green) of the system can be observed although a unique stable equilibrium (gray dashed line) is given for all values of $\phi$: (i) the green trajectory tracks the pathway of the moving stable equilibrium downwards the gray dashed line and (ii) the red trajectory undergoes a rate-induced critical transition resulting in a temporary collapse of the prey population. (B): Time series of the prey density $u$ when the system follows the moving stable equilibrium and (C): Time series of the prey density $u$ when the system undergoes a rate-induced critical transition.}
\label{fig:motivation}
\end{figure}

Critical transitions between alternative stable states have been a major focus of research in ecology during the last decades. Such transitions have been classified as continuous or abrupt under variation of external environmental conditions \citep{Scheffer2001, Scheffer2003}. While such critical transitions, from one stable ecosystem state to another, are called 'regime shifts' in ecology \citep{Scheffer2009}, similar transitions have been named 'tipping phenomena' in climate science \citep{Lenton2008,Lenton2013}. 
Although regime shifts have been first discussed in theoretical models \citep{Holling1973,May1977}, these critical transitions have been observed in several ecosystems in nature \citep{Hempson2018,Adam2011,Kosten2012,Folke2004,Steele1996,Gunderson2001,Foley2003,Scheffer2003}. The large interest in analyzing and predicting regime shifts is related to the fact that 
they can cause significant changes in the affected ecosystem such as population collapses \citep{Agler1999,Pinsky2011}, dominance changes in communities \citep{Scheffer1997,vanNes2007_2,Dudgeon2010} or even the extinction of species \citep{McCook1999,Aberhan2015}. Hence, their identification and early detection in ecosystems is relevant for the development of suitable management strategies to prevent future undesirable regime shifts \citep{Scheffer2009,Suding2004}. \\
Most of the work devoted to the study of critical transitions in ecology considers transitions
that occur due to the loss of stability via a classical bifurcation at some critical threshold of environmental conditions (e.g. critical resource concentration, atmospheric temperature, CO2-concentration) and often involve hysteresis phenomena \citep{Claussen2013,Faassen2015}. Linear stability analysis is employed to find these critical thresholds by constructing bifurcation diagrams reflecting stable and unstable ecosystem states, such as equilibria or limit cycles, for different but fixed-in-time environmental conditions \citep{Scheffer2003,vanNes2007,Scheffer2009,Scheffer2012,Bathiany2018}. Linear stability analysis only considers  
(i) small perturbations around the attractor to justify the linearization, and (ii) 
 time-independent (quasistatic) parameters which neglects changes of environmental conditions that might occur on the natural ecosystem timescales. Taking these two restrictions into account, one can identify 
critical parameter thresholds at which an attractor disappears or loses stability and a regime shift or tipping occurs. However, variations of environmental conditions which are comparable to or even faster than the internal ecosystem dynamics can be present \citep{Kees2011} and may even occur at unprecedented rates \citep{Joos2008}. Most importantly, environmental conditions can change at a rate at which the ecosystem is unable to adapt its behavior \citep{Walther2002}. Those environmental variations can lead to unfamiliar and often unexpected critical transitions, called \textit{rate-induced critical transitions}, that cannot be explained by linear stability analysis \citep{Luke2010,Wieczorek2011,Ashwin2012,Perryman2014,Siteur2016,OKeeffe2019}.\\
In figure \ref{fig:motivation}, we demonstrate such a rate-induced critical transition for the well-known Rosenzweig-MacArthur predator-prey model discussed in detail in the next section. The three-dimensional model consists of a fast-evolving prey population $u$, a slowly-reproducing predator population $v$, and slowly-varying environmental conditions $\phi$ at a constant rate $r>0$. For all fixed values of $\phi$, there exists a stable equilibrium where predator and prey coexist, meaning that there is no critical threshold of environmental conditions, i.e. no bifurcation occurs within the chosen range of $\phi$. 
In other words, we do not consider transitions to oscillatory solutions or the \mbox{\textit{paradox of enrichment}} \mbox{\cite{Rosenzweig1971}}.
However, when the environment changes  (the parameter $\phi$ changes over time), the position of the stable equilibrium in the $(u,v,\phi)$ phase space changes - it moves along the gray dashed line towards lower predator population densities.
The red and green trajectories in figure \ref{fig:motivation} demonstrate
that something unexpected and potentially catastrophic may happen in response
to slow environmental changes even though the moving equilibrium remains linearly stable. Both, the green and red trajectories start at the same initial state (the stable equilibrium), are subject to an environmental change that occurs at very similar rates $r$, but evolve drastically differently in time. The green trajectory tracks the moving stable equilibrium (gray dashed line) where predator and prey coexist as one would expect. However,
the red trajectory that is exposed to only slightly faster changing external conditions
shows a large deviation from the pathway of the moving stable equilibrium, resulting in a temporary collapse of the prey population to very low population densities.\\
While \mbox{$u$} remains low, small additional disturbances such as 
demographic or environmental noise could cause the extinction of the small prey population leading to an irreversible breakdown of the whole ecosystem \mbox{\citep{Liephold2003}. It is important to note that
the} unexpected temporary population collapse does not involve 
any classical bifurcations or any loss of linear stability. By contrast, it is induced entirely by the change of $\phi$ over time. In other words, this system possesses a critical rate of environmental change.\\
What is more, the critical rate is slower than the slowest timescale in the predator-prey system. 
This demonstrates another counter-intuitive aspect of rate-induced critical transitions: they may occur for
environmental changes that happen on a much slower timescale than the intrinsic ecological dynamics.
An in-depth analysis of the rate-induced critical transition shown in figure {\ref{fig:motivation}},
and the explanation of the underlying mechanism constitute the main focus of this paper.

The theoretical framework for analyzing rate-induced critical transitions in slow-fast systems has been developed by~\cite{Wieczorek2011}, where they apply their theory to a simple three-dimensional peatland soil model. They demonstrate that this model undergoes a rate-induced critical transition - a sudden release of soil carbon into the atmosphere - when the atmospheric temperature raises faster than some critical rate of environmental change. 

\cite{Morris2002},~\cite{Scheffer2008} and~\cite{Siteur2016} were 
the first to identify ecological models that are sensitive to the 
rate of environmental change, and to analyze rate-induced critical transitions 
in ecosystems. For example,~\cite{Siteur2016} studied the time-scaled Rosenzweig-MacArthur predator-prey model with time-dependent growth rate and demonstrated that this model undergoes a rate-induced critical transition of the same type as identified in \cite{Wieczorek2011} when the growth rate of the prey exceeds some critical rate of environmental change. To estimate the critical rate, they derived a \textit{steady-lag approach}. 
However, their method is only valid when 
the time lag between the system and the moving equilibrium is independent of the changing parameter, meaning that it does not apply for the time-scaled Rosenzweig-MacArthur predator-prey model with time-varying environmental conditions in general.\\
\nocite{Sarkar2018}
In this work, we close this gap and go beyond the first study by \cite{Siteur2016}.
We investigate again the three-dimensional time-scaled Rosenzweig-MacArthur predator-prey model with time-dependent parameter with respect to rate-induced critical transitions by employing a method from singular perturbation theory \citep{Fenichel1979} called \textit{desingularization} \citep{Dum1996,Krupa2001}. This method links rate-induced critical transitions in slow-fast systems to certain folded singularities,
and transforms the predator-prey model into a new system in which folded singularities become regular equilibria and can be analyzed using standard techniques from dynamical systems theory such as linear stability analysis \citep{Wieczorek2011,Wechselberger2013}. 
Studying the desingularized system reveals that a special solution called \mbox{\textit{maximal canard trajectory}} is the threshold separating tracking (green trajectory in fig. {\ref{fig:motivation}}) from rate-induced tipping (red trajectory in fig. {\ref{fig:motivation}}). Owing to the presence of two slow variables $v$ and $\phi$, this canard trajectory is generic and exists for large parameter regions. Note that
canards can also occur in systems with one slow variable, where they 
exist only within very narrow parameter regions, but can nonetheless have a significant impact on 
the dynamics \mbox{\citep{Wechselberger2001, Wechselberger2013}}. For instance, in the two-dimensional time-scaled Rosenzweig-MacArthur model, a maximal canard 
marks the sudden transition (canard explosion) between small-amplitude oscillations and large-amplitude relaxation oscillations \mbox{\citep{Kooi2018,Rinaldi1992}} following a Hopf bifurcation \mbox{\citep{Poggiale2019}}.\\
In contrast to \cite{Siteur2016} where the growth rate of the prey population increases in time, we choose the carrying capacity of the prey population, an ecological accessible parameter, to decrease over time at a given rate $r$ whitin a bounded $\phi$-interval (S. Sakar and P.S. Dutta, unpublished manuscript). 
The carrying capacity represents the maximum population density that the environment can sustain which, in turn, is given by the availability of resources in the habitat. Hence, a reduction of the carrying capacity of the prey population can be interpreted as a decline of resources caused by habitat loss, habitat fragmentation, habitat degradation or even destruction \citep{Zanette2000,Fischer2007,Mortelliti2008}. As these processes become more frequent and more widespread due to the increased land-use change by the growing human population \citep{Kees2011} and due to the effects of global climate change \citep{Selwood2015}, there is a growing need to better 
understand the ecosystems sensitivity to the rate of declining resources.\\
In the next section, we introduce the time-scaled Rosenzweig-MacArthur predator-prey model with a fixed-in-time carrying capacity to illustrate its most relevant dynamical patterns to which we refer throughout the paper. In section 'Population collapse due to rate-induced transitions', we assume the carrying capacity of the prey population to be time dependent by adding a third equation to the two-dimensional time-scaled predator-prey system, which describes the decline of the carrying capacity over time. We demonstrate that the now three-dimensional time-scaled predator-prey model undergoes a rate-induced critical transition when the resources decline faster than some critical rate. Moreover, we analyze the mechanism of the observed rate-induced critical transition and its ecological consequences. In section 'The tipping threshold - a canard comes into play', we find that the occurrence of rate-induced critical transitions in the system depends additionally on the initial density of predator and prey, and determine explicitly the {\em threshold} separating initial states that lead to tracking from those that can lead to population collapse. This approach allows us to analyze the dependence of the rate-induced critical transitions on additional model parameters such as the rate of environmental change itself. Finally, we discuss the generality of our results and their consequences for ecological systems.
\section{The slow-fast Rosenzweig-MacArthur predator-prey model}
\label{sec_model}
\begin{figure}[H]
\centering
\includegraphics[scale=0.6]{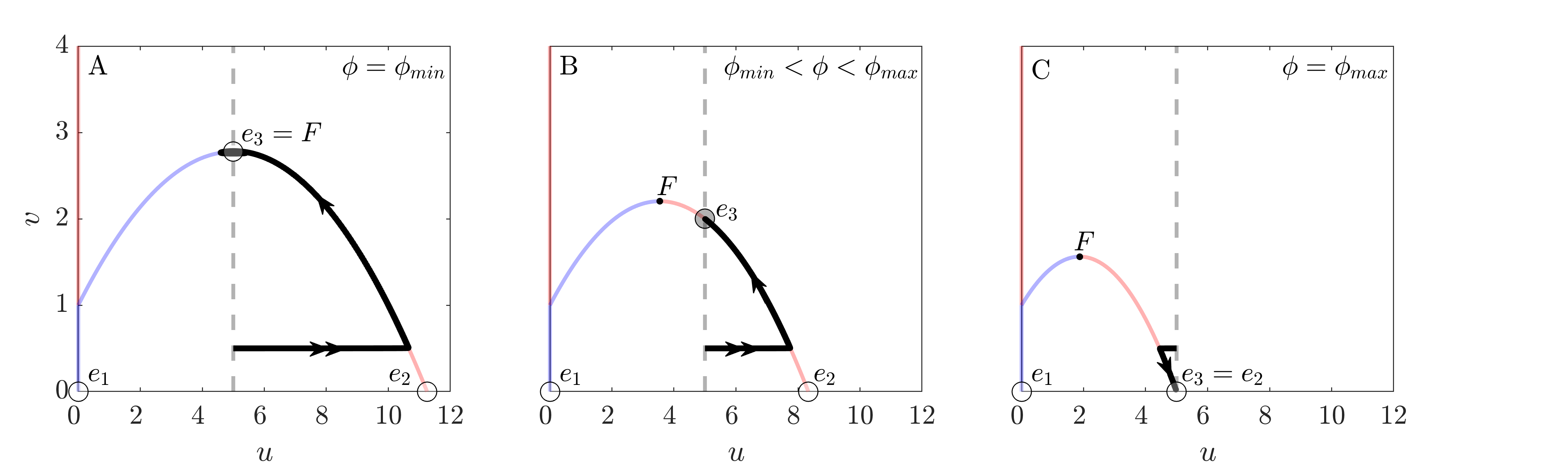}
\caption{Three phase portraits of the slow-fast Rosenzweig-MacArthur predator-prey model \mbox{\eqref{eq:final}}--\mbox{\eqref{eq:final2}} for $\phi$ increasing from $\phi_{min}$ to $\phi_{max}$ (A to C). Stable parts (red), unstable parts (blue) of the critical manifold $S_0$, its fold $F$ (black dot), unstable and marginally stable (open circles) and stable equilibria $e_1$, $e_2$ and $e_3$ (filled circles) and the predator's nullcline $v=(1-\eta)^{-1}$ (gray dashed line) are added to the phase portraits. Single arrows indicate slow motion and double arrows indicate fast motion. (A): The third equilibrium $e_3$ undergoes a Hopf-bifurcation. (B): Stationary coexistence of predator and prey and (C): The third $e_3$ and second equilibrium $e_2$ meet in a transcritical bifurcation. Parameters: $\kappa = 0.01$, $\eta=0.8$, $u_0 = 5$, $v_0 = 0.5$, $\phi = \phi_{min}=\frac{\eta(1-\eta)}{1+\eta}$ (A), $\phi=0.12$ (B) and $\phi = \phi_{max} = 1-\eta$ (C).}
\label{fig:RM_basic_dynamics_fastslow}
\end{figure}

The Rosenzweig-MacArthur predator-prey model is one of the most discussed paradigms in ecology \citep{Rosenzweig1963,Berryman1992}. It describes the biological interactions between predators $y$ and their prey $x$ using the logistic growth of the prey, Holling Type-II functional response, and linear mortality of the predator: 
\begin{align}
\label{eq:org}
&\frac{dx}{d\tau} = ax\left(1-\frac{x}{K}\right) - \frac{\alpha x y}{1+\beta x}\\
\label{eq:org2}
&\frac{dy}{d\tau} = \gamma \frac{\alpha x y }{1+\beta x} - cy.
\end{align}
Logistic growth is determined by the maximum per-capita growth rate of the prey $a$ and the carrying capacity $K$ which determines the maximum population density of the prey in the equilibrium when the predator is absent. The functional response of the predator is characterized by its maximum predation rate $\alpha/\beta$ and its half-saturation constant $1/ \beta$. Since not all prey taken up are converted into biomass of the predator, a conversion efficiency $\gamma$ is introduced to specify the ratio between biomass increase and food uptake. 
Predator mortality, represented by the term $-cy$, is assumed to be proportional to the predator population density.\\
In the following, we use a suitable coordinate transformation to obtain dimensionless variables and additionally to reduce the number of relevant parameters of the system. 
Specifically, we reformulate system~\eqref{eq:org}--\eqref{eq:org2} in terms of new variables 
$u=\alpha \gamma x/c$, $v=\alpha y/a$ 
and $t=\tau a$ to obtain the time-scaled Rosenzweig-MacArthur predator-prey model (see appendix \ref{app_A} or \mbox{\cite{Bazykin1998}} for more details):
%
\begin{align}
\label{eq:final}
\kappa\frac{du}{dt} &= u(1-\phi u) - \frac{uv}{1+\eta u}\\
\label{eq:final2}
\frac{dv}{dt} &= \frac{uv}{1+\eta u} -v 
\end{align}
where $\phi = c/(\alpha \gamma K)$ is proportional to the inverse of the carrying capacity $K$. The parameter $\eta = c \beta/(\alpha \gamma)$ can be interpreted as the predator's handling time of the prey, while $\kappa = c/a$ 
quantifies the timescale separation between prey's and predator's lifetime. 
In nature, predator and prey often correspond to different trophic levels at which they possess different lifetimes e.g. the lifetime of insects (prey) is much shorter as the lifetime of birds (predator). In the following we assume that the death rate of the predator $c$ is much smaller than the growth rate of the prey $a$, leading to a predator-prey system with a fast evolving prey and a slower reproducing predator population. In other words, $0<\kappa\ll 1$ becomes a small parameter. In addition, the predator's grazing rate $\alpha$ has to be high to compensate for the low conversion efficiency $\gamma$ \citep{Ginzburg1998}.\\
In general, the dynamics of the time-scaled system \eqref{eq:final}--\eqref{eq:final2} can be described as slow-fast because it consists of long periods of slow (single arrows in fig. {\ref{fig:RM_basic_dynamics_fastslow}}) that are occasionally interrupted by short episodes of fast change (double arrows in fig.  {\ref{fig:RM_basic_dynamics_fastslow}}) ~\citep{Fenichel1979,Cortez2010}. 
The slow motion can be approximated by the so-called \textit{critical manifold} \citep{Fenichel1979,Wechselberger2001,Cortez2010,Hek2010}.
More precisely, when $0<\kappa\ll 1$, the slow motion of the fast-slow system  \eqref{eq:final}--\eqref{eq:final2} 
takes place near the one-dimensional critical manifold 
\begin{equation}
\label{manifold2_1d}
S_0(\phi,\eta) = \left\lbrace(u,v) \in \mathbb{R}^2 : u\left((1-\phi u) - \frac{v}{1+\eta u}\right)=0, u \geq 0,v \geq 0\right\rbrace,
\end{equation} 
obtained by setting $\kappa=0$ in Eq.~\eqref{eq:final}. The
critical manifold $S_0$ consists of two components, the straight component $u=0$ and the folded component $v = (1-\phi u)(1+\eta u)$. The folded component has a fold
tangent to the fast $u$-direction at the point
$$
F(\phi,\eta) = (u_F,v_F) = \left(\frac{\eta-\phi}{2\eta\phi },\frac{(\eta + \phi)^2}{4\eta\phi} \right), 
$$ 
where the slow-motion approximation breaks down, and the system switches between slow 
and fast motion (see appendix~\ref{app_B} for more details). 
The two components of the critical manifold \mbox{$S_0$} \mbox{\eqref{manifold2_1d}} can be further divided into stable (red in figure {\ref{fig:RM_basic_dynamics_fastslow}}) and unstable  (blue in figure {\ref{fig:RM_basic_dynamics_fastslow}}) parts. Stable parts attract `fast' trajectories whereas unstable parts repel them.\\
%
In addition to the stable and unstable parts of the critical manifold \mbox{$S_0$} {\eqref{manifold2_1d}}, the dynamics of the time-scaled predator-prey system {\eqref{eq:final}}--{\eqref{eq:final2}} is determined by the stability and positions of the three equilibria \mbox{$e_1$}, \mbox{$e_2$} and \mbox{$e_3$}:
\begin{align}
e_1 &= (0,0),\\
e_2 &= \left(\frac{1}{\phi},0\right),\\
\label{e_3}
e_3 &= \left(\frac{1}{1-\eta},\frac{1-\eta -\phi}{(1-\eta)^2}\right).
\end{align}
Figure \ref{fig:RM_basic_dynamics_fastslow} visualizes the dynamics of the time-scaled predator-prey system  \eqref{eq:final}--\eqref{eq:final2} for fixed $\eta=0.8$ and different values of \mbox{$\phi$} increasing from \mbox{$\phi_{min}$} to \mbox{$\phi_{max}$} (fig. {\ref{fig:RM_basic_dynamics_fastslow}}A to fig.  {\ref{fig:RM_basic_dynamics_fastslow}}C). 
The first equilibrium $e_1$ is always located at the origin $(u,v) = (0,0)$ and is always unstable (open circle in fig. \ref{fig:RM_basic_dynamics_fastslow}A--\ref{fig:RM_basic_dynamics_fastslow}C). It represents the situation when predator and prey are extinct. The location of the second equilibrium $e_2$ is given by the intersection point of the critical manifold $S_0$ and the $u$-axis. It can be stable (not shown) or unstable (open circle in fig. \ref{fig:RM_basic_dynamics_fastslow}A--\ref{fig:RM_basic_dynamics_fastslow}C) depending on the parameters $\phi$ and $\eta$. When the equilibrium $e_2$ is stable, the prey grows to its carrying capacity while the predator dies out. The third equilibrium $e_3$ is given by the intersection point of the critical manifold $S_0$ and the predator nullcline $v=(1-\eta)^{-1}$ (gray dashed line). The stability of $e_3$ also depends on $\phi$ and $\eta$. A stable equilibrium $e_3$, as shown fig. \ref{fig:RM_basic_dynamics_fastslow}B, corresponds to a stable stationary coexistence of predator and prey. If $e_3$ is unstable, predator and prey coexist in oscillations (not shown). 
Figure \ref{fig:RM_basic_dynamics_fastslow}A  and \ref{fig:RM_basic_dynamics_fastslow}C  depict the two bifurcation situations where the third equilibrium $e_3$ is marginally stable. 
In figure \ref{fig:RM_basic_dynamics_fastslow}A for $\phi=\phi_{min}=\eta(1-\eta)/(1+\eta)$, the third equilibrium is located at the fold $F$ where it undergoes a Hopf-bifurcation, whereas in figure \ref{fig:RM_basic_dynamics_fastslow}C for $\phi=\phi_{max}=1-\eta$ the two equilibria $e_2$ and $e_3$ meet in a transcritical bifurcation where they exchange their stability. In figure \ref{fig:RM_basic_dynamics_fastslow}B, predator and prey coexist in the stable equilibrium \mbox{$e_3$} (filled circle). In the following, we study the system as shown in figure {\ref{fig:RM_basic_dynamics_fastslow}}B by limiting the change of \mbox{$\phi$} to \mbox{$\phi_{min}<\phi < \phi_{max}$}. Consequently, the third equilibrium $e_3$ is always stable and does not bifurcate. Hence, we exclude the situations where predator and prey coexist in oscillations and the prey grows to its carrying capacity because predators are extinct.

\section{Population collapse due to a rate-induced critical transition}
\label{sec_rate_induced_transitions}

\begin{figure}[H]
\centering
\includegraphics[scale=0.7]{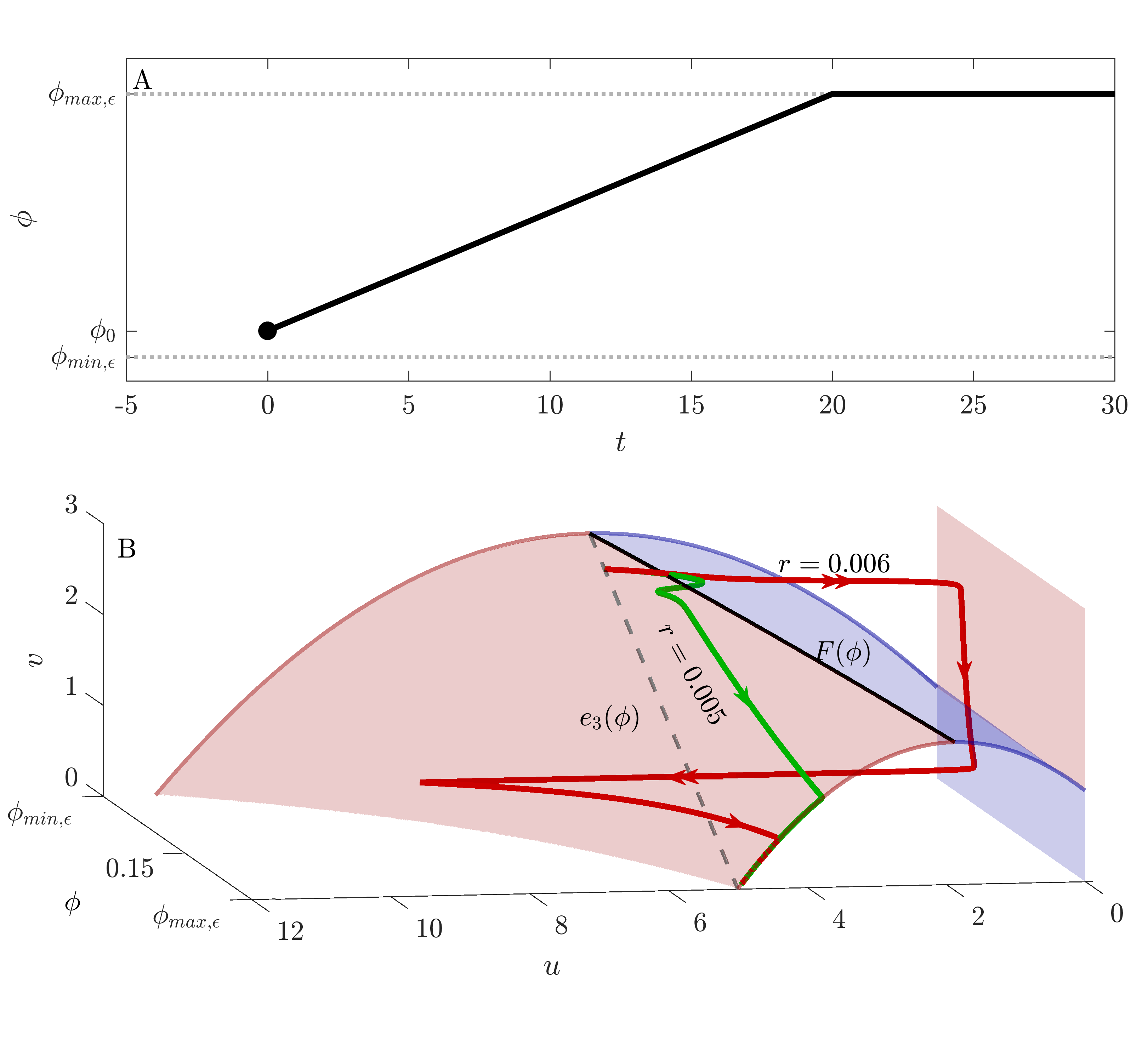}
\caption{(A): The parameter $\phi$ increases linearly in time $t$ from $\phi_0=\phi(0)>\phi_{min,\epsilon}$ to $\phi_{max,\epsilon}$ at the rate $r$. (B): Phase portrait of the ramped system \eqref{rampedRM1}--\eqref{phi} for linearly decreasing resource concentration $\phi$ at the rates $r=0.005$ (green) and $r=0.006$ (red). Stable (red) and unstable parts (blue) of the critical manifold $S_0$, the fold $F(\phi)$ (black solid line) and the pathway of the moving stable equilibrium $e_3(\phi)$ (gray dashed line) are added to the phase portrait. When $r=0.005$ (green trajectory), the system tracks the moving equilibrium $e_3(\phi)$. When $r=0.006$ (red trajectory), the system undergoes a rate-induced critical transition leading to a temporary collapse of the prey population. Parameters: $\phi_0 = 0.1$, $r=0.006$, \mbox{$\epsilon = 10^{-6}$} (A) $\kappa = 0.01$, $\eta = 0.8$, $\phi_0 = 0.1$, $u_0 = (1-\eta)^{-1}$, $v_0 = (1-\phi_0 u_0)(1+\eta u_0)$ (B).}
\label{fig:ramping_process}
\end{figure}

According to our aim to study rate-induced critical transitions, we relax the assumption that environmental conditions remain constant until the system reaches its stable equilibrium, and expose the prey population to a continuous decline of resources within its habitat. This could be e.g. due to a land-use change leading to less food for the prey population or due to an increasing fragmentation of the landscape which would cut the access to certain resources. 
Specifically, we model the decline of resources by starting from $\phi_{min,\epsilon}=\phi_{min}+\epsilon$ and increasing the parameter \mbox{$\phi$} over time at a constant rate \mbox{$r$} until it reaches the maximum value $\phi_{max,\epsilon}=\phi_{max}-\epsilon$ for some \mbox{$0<\epsilon \ll 1$} (see fig. {\ref{fig:ramping_process}}A). 
This ensures that equilibrium $e_3$ remains stable for all values of \mbox{$\phi$}.
Notice that increasing $\phi$ corresponds to lowering the carrying capacity $K$ and therefore to reducing the available resources within the habitat.

The dimensionless Rosenzweig-MacArthur model with a continuous decline of the prey's resources $\phi$ is given by the following \textit{three} equations, including the additional equation for the time evolution of $\phi$:

\begin{align}
\label{rampedRM1}
\kappa \frac{du}{dt} &= u(1-\phi(t) u) - \frac{uv}{1+\eta u} \\
\label{rampedRM2}
\frac{dv}{dt} &= \frac{uv}{1+\eta u} - v \\
\label{phi}
\frac{d\phi}{dt} &= \begin{cases}
r & \text{if}\;\; \phi_{min,\epsilon} < \phi < \phi_{max,\epsilon}\\
0 & \text{otherwise},
\end{cases}
\end{align}

A comparison between the position of the equilibrium $e_3$ in figures \ref{fig:RM_basic_dynamics_fastslow}A and \ref{fig:RM_basic_dynamics_fastslow}C shows that its position changes to lower predator densities along the predator's nullcline when $\phi$ is increasing in time because the folded component of the critical manifold $S_0$ shrinks. Hence, we write for the moving stable equilibrium $e_3=e_3(\phi)$ when $r\neq0$.  
Since the linear increase of $\phi(t)$ resembles a ramp, we 
refer to the three-dimensional system~\eqref{rampedRM1}--\eqref{phi} 
as the \textit{ramped system}. The ramped system evolves in the three-dimensional
$(u,v,\phi)$ phase space where $\phi(t)$ becomes the second slow variable.
Thus, the two components of the slow manifold $S_0$ become a two-dimensional plane located at $u = 0$ and a two-dimensional surface folded along the curve $F(\phi)$ as shown in figure~\ref{fig:ramping_process}B.
When \mbox{$\phi$} increases linearly in time at the rate $r$, the moving stable equilibrium \mbox{$e_3(\phi)$} proceeds downwards the grey dashed line (predator's nullcline) causing the predator population density $e_{3,y}$ to shrink whereas the prey population density $e_{3,x}$ stays constant. 
Hence, we conclude that lowering slowly the resources in the habitat results in a stabilization of the prey population and in an increased threat to the predator population which is close to extinction at \mbox{$\phi = \phi_{max,\epsilon}$}. Therefore, one would expect that lowering the prey's resources $\phi$ at a given rate $r$ leads to a stabilization of the prey population in the model. \\
In figure~\ref{fig:motivation}, we have already demonstrated that the stabilization of the prey population is just one possible solution of the ramped system \eqref{rampedRM1}--\eqref{phi} which is represented by the green trajectory that \textit{tracks} the moving stable equilibrium $e_3(\phi)$ (gray dashed line). The other possible dynamics is demonstrated by the red trajectory which exhibits a large deviation from the pathway of the moving stable equilibrium $e_3(\phi)$ and undergoes a \textit{rate-induced critical transition} leading to a temporary collapse of the prey population. In the following, we call this behavior of the ramped system \textit{rate-induced tipping} or simply {\em R-tipping} \citep{Ashwin2012}. To give a first explanation of the green and red trajectory's pathway in figure~\ref{fig:motivation} we employ concepts from the theory of fast-slow systems. To this end, we add the two-dimensional stable and unstable parts of the critical manifold $S_0$ to the green and red trajectories (see fig. \ref{fig:ramping_process}B).\\
Both, red and green trajectory, start at the moving stable equilibrium $e_3(\phi)$ (gray dashed line) on the stable part of the folded component (red) and proceed close to each other towards the fold $F(\phi)$. At the fold $F(\phi)$, they start to move quite differently: both cross the fold $F(\phi)$ but the red trajectory moves fast towards the stable part of the two-dimensional straight component of $S_0$ whereas the green trajectory reverses and dives under the folded component and returns to the stable part of the folded component. The explanation of these  different pathways is subject of the following section 'The tipping threshold - a canard comes into play'. Here, we focus on the further fate of both trajectories. On the stable part of the folded component of $S_0$, the green trajectory stays always close to the pathway of the moving equilibrium $e_3(\phi)$. Thus, the green trajectory tracks the moving equilibrium $e_3(\phi)$. When the red trajectory proceeds slowly down the stable part of the straight component of $S_0$, the prey population density remains very low ($\approx 10^{-12}$). Hence, a small perturbation or noise could lead to its extinction. When the straight component of $S_0$ turns unstable, the red trajectory is repelled towards the stable part of the folded component of $S_0$ and finally converges to the moving stable equilibrium $e_3(\phi)$ at \mbox{$\phi=\phi_{max,\epsilon}$}. Therefore, the ecological consequence of this rate-induced tipping is a possible
temporary \textit{collapse} of the prey population which may lead to its extinction when some noise is taken into account followed by the extinction of the predator and subsequently a breakdown of the entire ecosystem. 
The temporary collapse of the prey population is caused by overconsumption of the predator population. The prey population responds quickly to the decreasing resources by reducing its population density. By contrast, the density of the slower evolving predator population stays almost constant. As a result, a small prey population is confronted with a large predator population causing overconsumption and finally the temporary collapse of the prey population (see appendix {\ref{app_temporary_collapse}} for a more detailed analysis). In the ramped system \mbox{\eqref{rampedRM1}}--\mbox{\eqref{phi}}, the consumption of the prey population by the predators \mbox{$\frac{uv}{1+\eta u}$} depends on the time they need to kill, to hunt and to digest the prey which is given by the predator's handling time \mbox{$\eta$}. In appendix {\ref{app_impact_eta}}, we demonstrate that the occurrence of the rate-induced collapse crucially depends on the choice of this parameter.\\
Interestingly and rather surprisingly, this rate-induced critical transition occurs for environmental changes that are slower than the intrinsic timescales within the ecosystem (e.g. the prey growth rates or the predator mortality rates). 
According to \cite{Edwards1999}, most zooplankton-phytoplankton predator-prey models assume that the maximum growth rate of phytoplankton ranges between $a \in [1.4\; 1.75]$ day$^{-1}$ whereas the zooplanktons's mortality ranges between $c \in [0.015\; 0.15]$ day$^{-1}$ and, hence, $\kappa \in [0.01\; 0.08]$. Converting the non-dimensional rate $r=0.006$ into the dimensional time $\tau$ day$^{-1}$ leads to the fastest environmental changes at $d\phi/d\tau = r \kappa a = 1.05\cdot10^{-4}$ day$^{-1}$ with $\kappa = 0.01$ and $a=1.75$ which is much slower than the maximum growth rate of phytoplankton $a$ or the predator's mortality rate $c$.
To explain this counter-intuitive behavior, the intrinsic timescales in the ecosystem should be compared with the speed of the moving stable equilibrium 
$$
|\dot{e}_3| = \Big|\frac{d e_3}{dt}\Big| = \Big|\frac{d e_3}{d\phi}\Big|\,\frac{d\phi}{dt} = \frac{r}{(1-\eta)^2}\,,
$$
rather than with the rate $d\phi/dt = r$ of the environmental change alone~(\cite{Ashwin2012}). For a rate of environmental change of $r=0.006$, a reproduction rate $a=1.75$ day$^{-1}$ and a predator's handling time $\eta=0.8$, the equilibrium $e_3(\phi)$ moves at a speed of $|de_3/d\tau| = 0.0026$ day$^{-1}$ which is about one order of magnitude faster than the fastest environmental change $d\phi/d\tau = 1.05\cdot10^{-4}$ day$^{-1}$. 
In the ecosystem model, the position of the stable equilibrium $e_3(\phi)$ depends strongly on the parameter $\phi$. This causes the factor $de_3/d\phi=1/(1-\eta)^2$ to be large for $\eta$ sufficiently close to one, and gives a fast-moving stable equilibrium (or large $|\dot{e}_3|$) for  $d\phi/dt =r$ smaller than the intrinsic timescales. 

\newpage
\section{A tipping threshold - a canard comes into play}
\begin{figure}[H]
\centering
\includegraphics[scale=0.7]{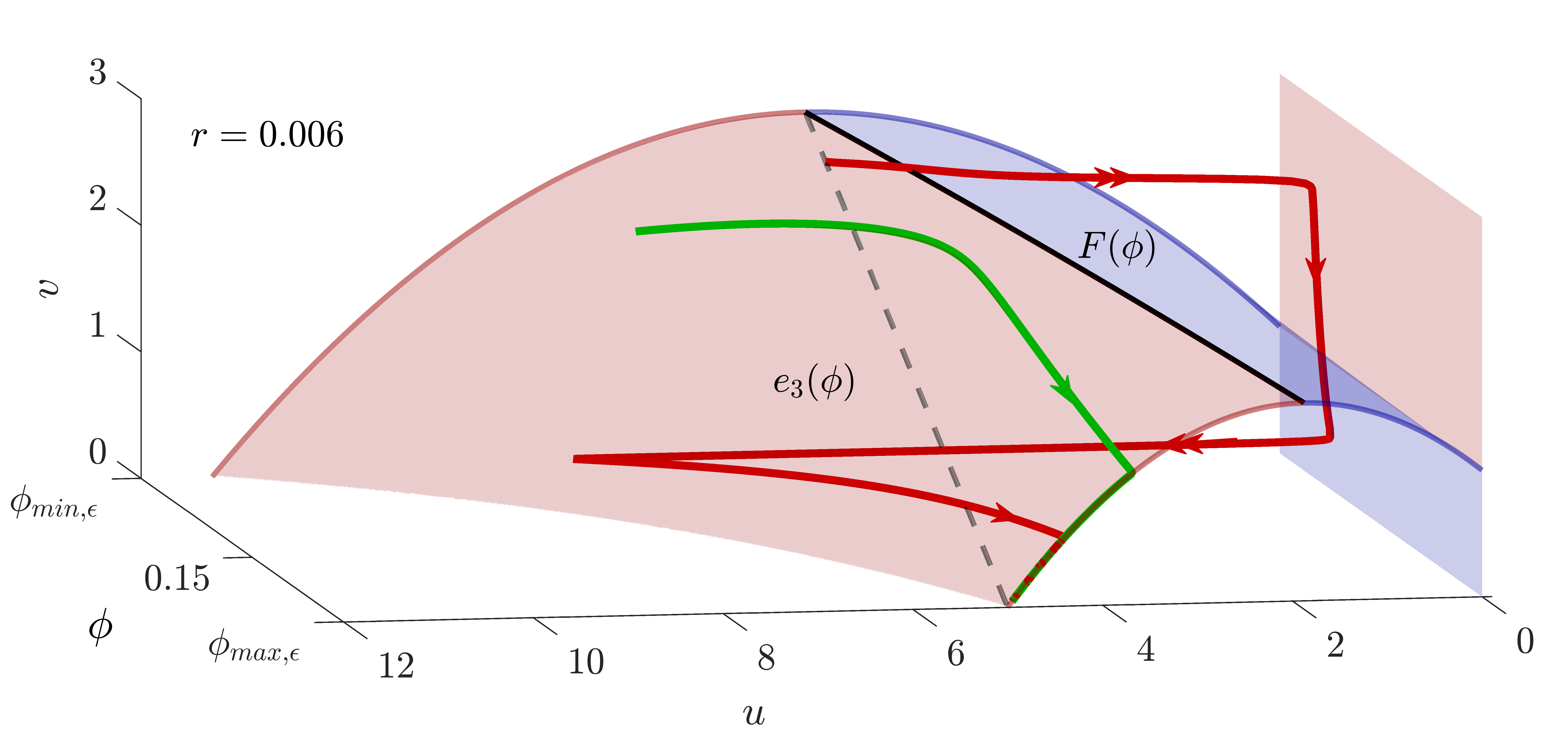}
\caption{Dependence of the rate-induced critical transition on the initial condition of the ramped system \eqref{rampedRM1}--\eqref{phi}, with two-dimensional critical manifold $S_0$ and its stable parts (red), unstable parts (blue), fold $F(\phi)$ (black solid line) and pathway of the moving stable equilibrium $e_3(\phi)$ (gray dashed line). 
The green trajectory tracks the moving stable equilibrium $e_3(\phi)$ while the red trajectory exhibits rate-induced tipping. Parameters: $\kappa = 0.01$, $\eta = 0.8$, $\phi_0 = 0.1$, \mbox{$\epsilon=10^{-6}$} (red): $u_0 = (1-\eta)^{-1}$, $v_0 = (1-\phi_0 u_0)(1+\eta u_0)$, (green): $u_0 = 7.0$, $v_0 = (1-\phi_0 u_0)(1+\eta u_0)$. }
\label{fig:dependence_on_ini}
\end{figure}

%
%
\label{folded_canard}
In the previous section, we fixed the initial state of the ramped system and 
demonstrated that its dynamics depends crucially
on the rate $r$. In particular, figure \ref{fig:ramping_process}B shows that the ramped system starting at the same initial state changes from tracking to rate-induced tipping at the critical rate  
$0.005<\hat{r}_{crit}<0.006$. In this section, we fix the rate $r$ and analyze the dynamics of the ramped system depending on the initial state. Therefore, we study the ramped system for two different initial conditions which are exposed to the same rate of environmental change $r=0.006$. Figure \ref{fig:dependence_on_ini} shows the (red) tipping trajectory from figure  \ref{fig:ramping_process}B together with an additional (green) 
trajectory that is started from a different initial state and which  tracks the moving stable equilibrium $e_3(\phi)$. Clearly, in addition to the rate $r$, the occurrence of rate-induced critical transitions in the ramped system depends on the initial density of predator $v_0$ and prey $u_0$ as well as the initial resource concentration $\phi_0$. Thus, a natural question emerges: Where is the boundary between the two initial states on the stable part of the folded component of $S_0$? More precisely, what is the {\em tipping threshold} that separates the initial states that track the moving stable equilibrium $e_3(\phi)$ from those that undergo a rate-induced critical transition at a fixed rate $r$?

\begin{figure}[H]
\centering
\includegraphics[scale=0.6]{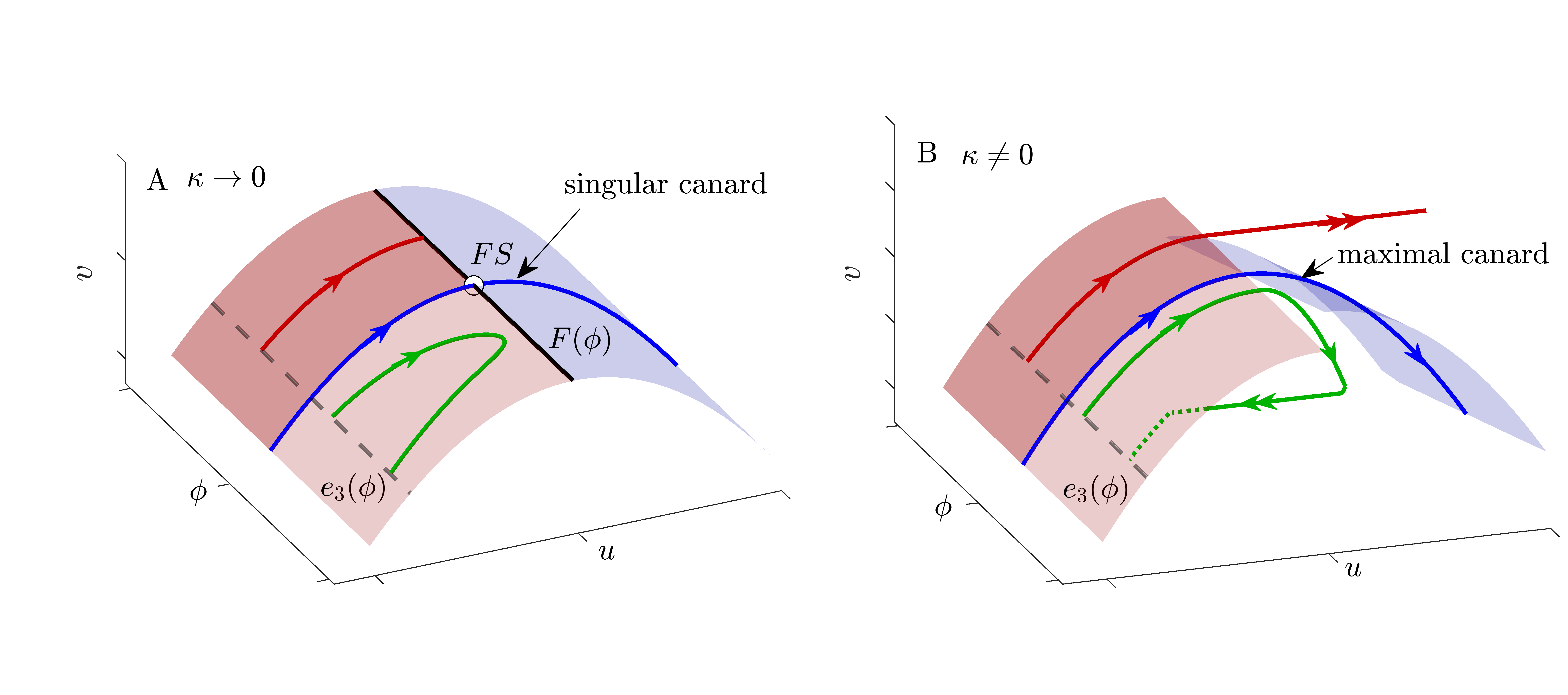}
\caption{(A): Sketch of the critical manifold $S_0$ of the ramped system \eqref{rampedRM1}--\eqref{phi} with its stable parts (red) and unstable parts (blue). The red trajectory cease to exist at the fold $F(\phi)$ (black solid line) whereas a singular canard (blue trajectory) is able to cross the fold $F(\phi)$ via the point $FS$ with finite speed. The green trajectory is repelled by the fold $F(\phi)$ and converges to the moving stable equilibrium $e_3(\phi)$ (gray dashed line). (B): When $\kappa \neq 0$, stable and unstable part of the critical manifold are displaced (perturbed) along the fold and are only intersecting at the point $FS$ at the fold. Via the intersection $FS$, a canard of the full system, called maximal canard, is able to proceed from the perturbed stable part (red) towards the perturbed unstable part (blue). Trajectories starting on the perturbed stable part above the maximal canard (dark red region) undergo a rate-induced critical transition whereas trajectories starting below the maximal canard (light red region) track the stable moving equilibrium $e_3(\phi)$.} 
\label{fig:folded_canard}
\end{figure}

To address this question we need to analyze the slow 
dynamics on the critical manifold $S_0$. These dynamics
are governed by the so-called \textit{reduced system} 
\begin{align}
\label{algebraic constrain}
0 &= u(1-\phi u) - \frac{uv}{1+\eta u},\\
\label{rs1}
\frac{dv}{dt} &= \frac{uv}{1+\eta u} - v,\\
\label{rs2}
 \frac{d\phi}{dt} &=r,
\end{align} 
that is obtained by setting $\kappa = 0$ in the ramped system, and which describes the 
slow-time evolution of $v$ and $\phi$.
However, rate-induced tipping occurs in the fast variable $u$, meaning that
we need to study the  evolution of $u$ in slow time $t$. To this end, we differentiate the critical manifold condition~\eqref{algebraic constrain} with respect to $t$, use Eqs.~\eqref{rs1}--\eqref{rs2}, and obtain the alternative reduced system:
\begin{align}
\label{fast_flow_in_slow_time}
\frac{du}{dt} &= 
\frac{u(1-\phi u) - (1-\phi u)(1 + \eta u) + u(1 + \eta u)r}{2\phi\eta\left( u_F - u\right)}:= \frac{\Lambda(u,\phi,\eta,r)}{2\phi\eta\left( u_F - u\right)}\\
\label{rs2b}
 \frac{d\phi}{dt} &=r.
\end{align}
The key observation is that the denominator of $du/dt$ becomes zero at the fold $F$, where $u = u_F$. 
This gives rise to three types of trajectories within the stable part of the critical manifold where $u_F-u>0$ (see fig. \ref{fig:folded_canard}A). 
Firstly, there are (red) trajectories with $\Lambda<0$ and $du/dt < 0$. Such trajectories are attracted to  the fold $F(\phi)$, where $du/dt$ becomes infinite and the fast variable $u$ goes to infinity in finite slow time (blows up).
In other words, the solution ceases to exist within $S_0$ when it reaches $F(\phi)$.
Secondly, there can be special points $FS$ along the fold $F(\phi)$ 
at which $\Lambda$ also goes through zero such that $du/dt$ remains finite. The corresponding (blue) trajectory crosses $F(\phi)$ with finite speed and continues along the unstable part of $S_0$. This special trajectory is called the {\em singular canard}, and the special point $FS$ found in the reduced system~\eqref{fast_flow_in_slow_time}--\eqref{rs2b} is called the {\em folded saddle singularity}\footnote{ There can be other types of folded singularities, e.g. folded node singularities, giving rise to more complicated thresholds~(\cite{Perryman2014}).}~(\cite{Wechselberger2001}).
Thirdly, as $\Lambda$ changes sign at $FS$, the fold $F(\phi)$ changes 
from attracting ($\Lambda<0$) to repelling ($\Lambda>0$). Thus, (green) 
trajectories starting on the other side of the singular canard never reach $F(\phi)$ because they are repelled from the fold and proceed towards the moving stable equilibrium $e_3(\phi)$ 
(gray dashed line).
%
%

Now, we need to translate the singular limit ($\kappa = 0$) dynamics to the ramped system~\eqref{rampedRM1}--\eqref{phi} with a finite timescale separation between 
the prey and predator lifetimes ($0<\kappa\ll 1$). When $0<\kappa\ll 1$, the stable 
and unstable parts of the folded component of the critical manifold perturb to nearby stable and unstable 
{\em slow manifolds} (see fig. \ref{fig:folded_canard}B). The slow manifolds disconnect along $F(\phi)$, except for
one point $FS$ where they intersect along the {\em maximal canard} trajectory
(blue trajectory in fig.~\ref{fig:folded_canard}B).
(Red) Trajectories starting in the dark red region above the maximal canard  
move towards $F(\phi)$, reach the boundary of the stable slow manifold, and jump off in the fast $u$-direction above the unstable part of the slow manifold. (Green) Trajectories starting in the light red region below the maximal canard also move towards $F(\phi)$ and reach the boundary of the stable slow manifold. The difference is that these trajectories find themselves underneath the unstable slow manifold, are repelled straight back towards the stable slow manifold, and 
proceed to tracking the moving stable equilibrium $e_3(\phi)$. Hence,
this maximal canard is the {\em tipping threshold} separating initial states on the stable part of the slow manifold that undergo a rate-induced critical transition (dark red region in fig. \ref{fig:folded_canard}B) from those that track 
the moving stable equilibrium $e_3(\phi)$ (light red region fig. \ref{fig:folded_canard}B). In the case of the folded saddle singularity \mbox{$FS$}, there exist a one-to-one correspondence between singular and maximal canard \mbox{\citep{Wechselberger2013}}. Hence, we compute the singular canard trajectory and include it in phase portrait of the ramped system \mbox{\eqref{rampedRM1}}--\mbox{\eqref{phi}} in figure {\ref{fig:dependence_on_ini_with_canard}} (see appendix {\ref{app_C}} for more details). As shown in figure {\ref{fig:dependence_on_ini_with_canard}}, the singular canard (blue trajectory) is the tipping threshold separating the (green) tracking from the (red) rate-induced tipping trajectory.
\\ 

%
%
%
\begin{figure}[H]
\centering
\includegraphics[scale=0.7]{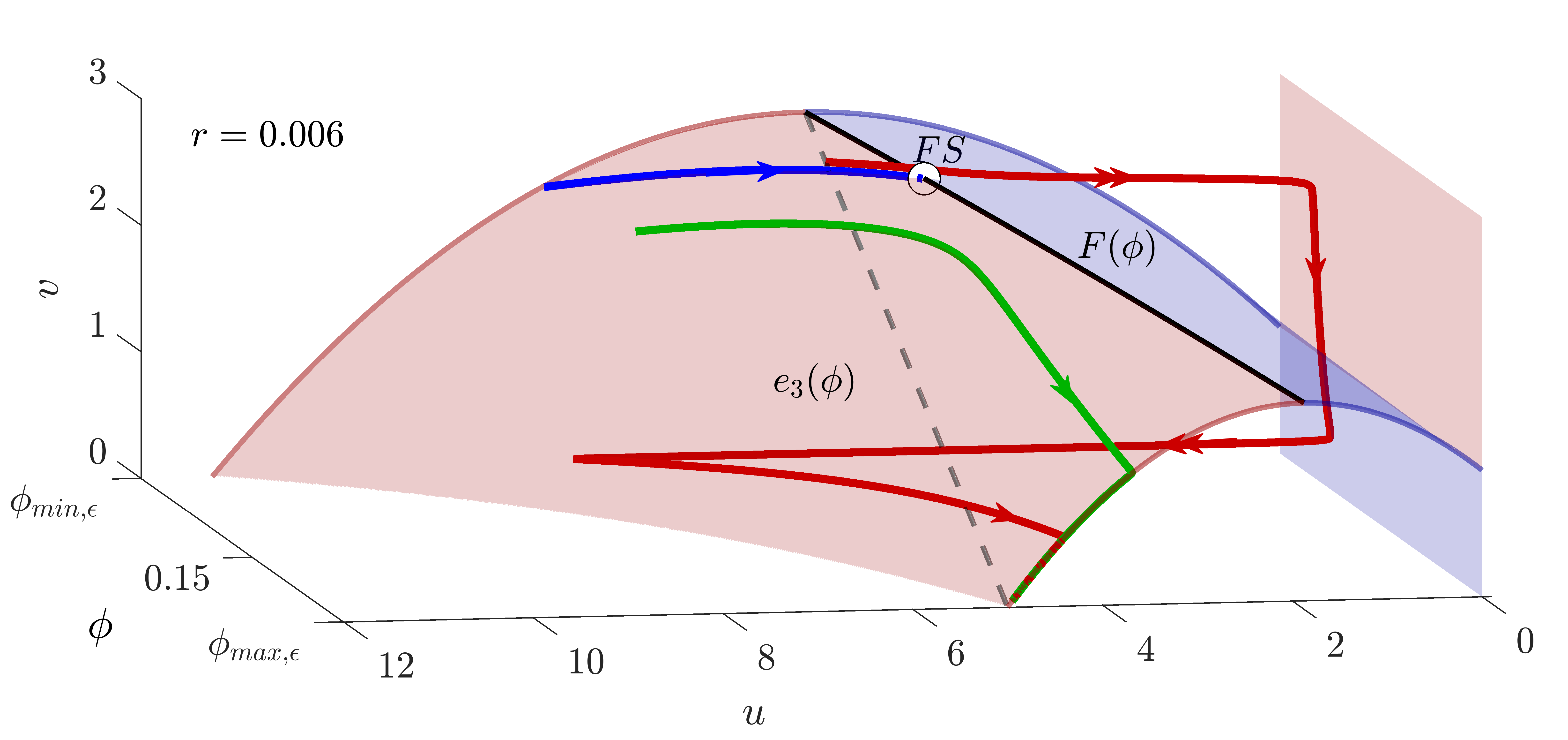}
\caption{Location of the tipping threshold (singular canard) and the corresponding folded saddle singularity $FS$ (white circle) between the tracking (green) and the tipping trajectory (red). All other notations and parameters as given in figure \ref{fig:dependence_on_ini}.}
\label{fig:dependence_on_ini_with_canard}
\end{figure}

\section{The impact of the rate of environmental change}
\label{sec_the_rate_of_change}

Let us now address the question of how the extent of the collapse-prone initial states (initial states above the singular canard) on the stable part of the folded component of the critical manifold $S_0$ changes when the rate $r$ of the environmental change increases. Figure \ref{fig:ramping_process}B demonstrates that the trajectories starting from a single initial state can switch from tracking to rate-induced tipping in response to a slight increase in $r$ from $r=0.005$ to $r=0.006$. 
We now extend this result to the whole set of initial states - we analyze how the position of the singular canard, which approximates  the tipping threshold within the stable part of the critical manifold, depends on the rate of environmental change $r$. 

\begin{figure}[H]
\centering
\includegraphics[scale=0.55]{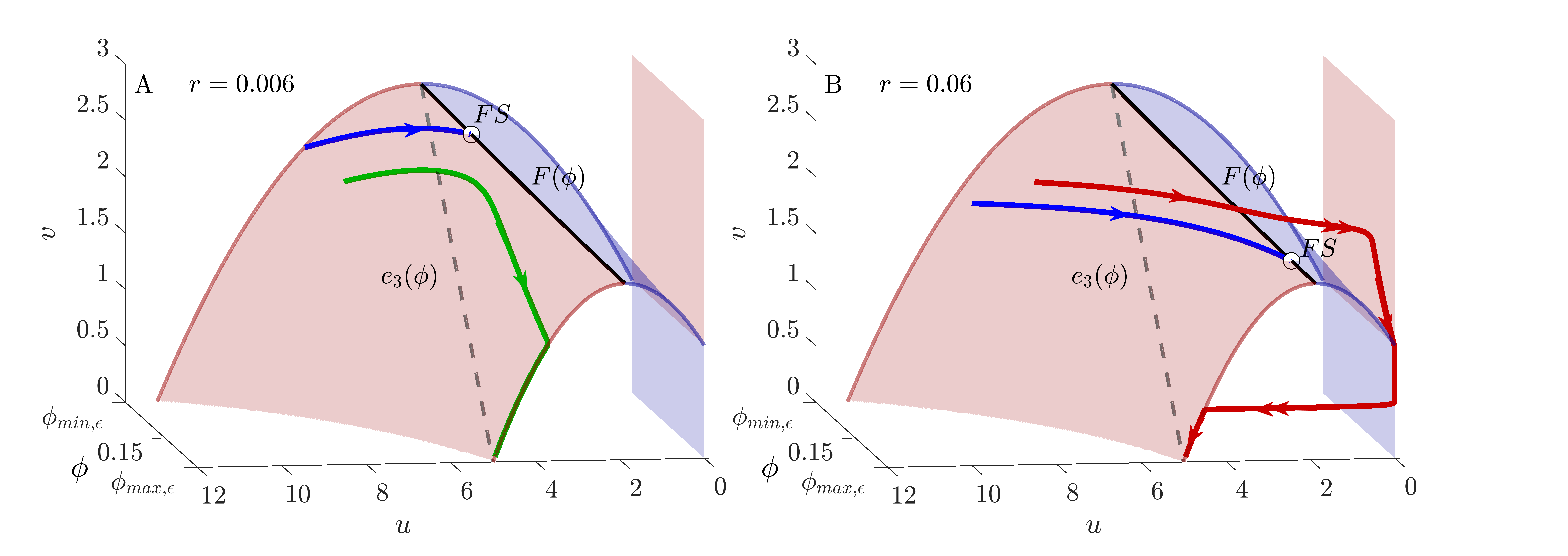}
\caption{Location of the tipping threshold (singular canard, blue line) and the corresponding folded saddle singularity $FS$ (white circle) depending on the rate of environmental change $r=0.006$ (A) and $r=0.06$ (B). The tipping threshold is located closer to $\phi_{max}$ the higher the rate of environmental change $r$. All other notations and parameters as in figure \ref{fig:dependence_on_ini}. }
\label{fig:affect_of_the_rate}
\end{figure}

Specifically, we compute the singular canard  for two different rates  $r=0.006$ (fig. \ref{fig:affect_of_the_rate}A) and  $r=0.06$ (fig. \ref{fig:affect_of_the_rate}B). For each rate $r$, we start the trajectory at the same initial state $(u_0,v_0,\phi_0)$ and use the same parameter values for $\kappa$ and $\eta$. 
When the system is exposed to the slower rate of environmental change $r=0.006$, the initial state is located `below' the singular canard 
threshold (blue line) within the stable part of the critical manifold. Thus, the (green) trajectory is repelled from the fold $F(\phi)$ and tracks the moving stable equilibrium $e_3(\phi)$. As the rate of environmental change is increased to $r=0.06$, the folded saddle equilibrium $FS$ (white circle) and the singular canard shift towards $\phi_{max}$ such that the fixed initial state is located 'above' the singular canard trajectory.
As a result, the (red) trajectory is attracted towards the fold $F(\phi)$ and exhibits the population collapse (see fig. \ref{fig:affect_of_the_rate}B).

In summary, an increase in the rate of environmental change $r$ shifts the position of the singular canard trajectory within the stable part of the critical manifold towards higher values of $\phi$. This leads to an expansion of the collapse-prone region of initial states located above the singular canard (see dark red region in fig. \ref{fig:folded_canard}A and \ref{fig:folded_canard}B). 
In other words, under variation of $r$, some initial states within the stable part of the critical manifold are crossed by the moving tipping threshold and become prone to rate-induced tipping.
Thus, initial states which are located closer to $\phi_{max}$ are less prone to rate-induced critical transitions because they are crossed by the tipping threshold at higher rates of environmental change. 


\section{Discussion}
\label{sec_discussion}
In the face of global climate change, rates of environmental change exceed those which were recorded in the past \citep{Smith2015}. Hence, some ecosystems are not able to adapt resulting in unexpected transitions of their system state.
Such transitions which are rather triggered by rates of environmental change than by its magnitudes are called rate-induced critical transitions. They can occur very unexpected even if a unique stable equilibrium exists for all values of the varying environmental conditions. Further, they can not be studied by the most commonly used methods in ecology: linear stability analysis and bifurcation theory because both methods assume a constant environment until the system is in equilibrium.\\
We have studied rate-induced critical transitions in the well-known Rosenzweig-MacArthur predator-prey system by using techniques of fast-slow systems theory instead of steady-state analysis. We showed that the predator-prey system, consisting of a fast evolving prey population and a much slower reproducing predator population, can undergo such an unexpected rate-induced critical transition when the carrying capacity of the prey population reduces with a given rate of environmental change. A similar transition has been found earlier by \cite{Siteur2016} changing the growth rate of the prey in time. This rate-induced critical transition can lead to a breakdown of the prey population due to an overconsumption by the predators when perturbations and/or noise would be taken into account. This leads in the end to a breakdown of the whole ecosystem. This clearly emphasizes the huge relevance of taking time-dependent environmental conditions into account when evaluating the stability and dynamics of ecosystems. In addition, rate-induced critical transitions in this simple ecosystem model can already occur at rates of environmental change which are much slower than its intrinsic dynamics. This reinforces the impression that we might miss a lot of possible transitions within ecosystems when we assume a time-independent environment.\\
 Furthermore, we have shown that the occurrence of rate-induced transitions in the system also depends crucially on its initial state. We are able to identify all initial states of the system which are prone to a rate-induced transition at a given rate by determining the collapse-threshold on the critical manifold. The collapse-threshold is given by an invariant set called the singular or maximal canard trajectory. Hence, we can deduce the biological characteristics of the collapse-prone systems which is useful for designing e.g. suitable protection measures. We have evaluated that predator-prey systems determined by a large inefficient predator population already exhibit rate-induced tipping at very small rates of environmental change. Nevertheless, the number of collapse-prone initial states increases when the environmental changes are accelerating. As a consequence also predator-prey systems with smaller but more efficient predator populations undergo rate-induced critical transitions.\\
\mbox{\cite{Poggiale2019}} studied canard trajectories in the two--dimensional Rosenzweig-MacArthur predator-prey model with one slow and one fast variable, showed that canards exist only in very small parameter ranges and concluded that they are of limited importance for predator-prey systems in general. However, this is true only 
for systems with one slow variable. In systems with at least two slow variables, such as the ramped Rosenzweig-MacArthur predator-prey model analyzed here, canard trajectories are generic and exist within large parameter intervals \mbox{\citep{Wechselberger2001}}. What is more, they have an important meaning of non-obvious R-tipping thresholds \mbox{\citep{Wieczorek2011,Perryman2014}}. Hence, our analysis goes beyond the results of \mbox{\citep{Poggiale2019}} and emphasizes the importance of canards for rate-induced critical transitions in ramped predator-prey systems possessing folded critical manifolds and timescale separation.
\\
In the Rosenzweig-MacArthur model, predator and prey population can only respond to environmental changes by adjusting their population densities. However in nature, populations are able to respond in other ways to rapid environmental change which can preserve them from extinction (e.g. due to a rate-induced critical transition). Some studies outline that populations migrate to more suitable habitats when they can not cope with certain rates of environmental change. For example the distribution of birds and butterfly populations shifts northwards due to an accelerating warming of their original habitats \citep{Parmesan1999,Thomas1999}. Therefore, the incorporation of space into the predator-prey system might prevent the collapse of the prey population. But even if populations are not able to migrate due to barriers such as mountains, lakes or streets, or just because they are not mobile (e.g. plants), they might be able to  adapt evolutionary to rapid environmental change at ecological time scales \citep{Yoshida2003, Jump2005, Sih2013}. This rapid evolutionary adaptation can be modeled e.g. with the quantitative genetic approach by \cite{Lande1982} and \cite{Abrams1993} where the mean trait value of a population changes at a rate proportional to the additive genetic variance of a trait and the individual fitness gradient \citep{Cortez2010}. Some studies suggest that most of the  adaptation involves phenotypic plasticity rather than immediate genetic evolution \citep{Sih2010,Sih2013}.  \\
Apart from the absence of biological mechanisms that might prevent the system from tipping, the identification of initial conditions that undergo a rate-induced critical transition depends crucially on the computation of the singular respectively the maximal canard trajectory. In the Rosenzweig-MacArthur model, their  explicit computation is only possible (i) due to the very simple model equations and (ii) because we have chosen the simplest approach to change the external environmental conditions by just ramping the model parameter $\phi$ linearly in time. When the model formulation or the change of the environmental condition in time or even both become more complicated, as for example for ecosystems with spatially periodic patterns where rate-dependent behavior is suggested \citep{Siteur2016}, it is challenging to compute the collapse-threshold and the associated critical rate of environmental change. Methods which enable the computation of the singular canard in more complex models would be useful to study rate-induced critical transitions in such ecosystems. In general, we expect rate-induced critical transitions to occur in other population dynamical models because many of them are determined by different time scales in the life cycles of populations on different trophic levels \citep{West2005} and non-linear terms describing the interactions of the species \citep{Hastings1997} that lead to  folded critical manifolds which is the main condition for the occurrence of rate-induced critical transitions.\\ 
Although rate-induced critical transitions are not reported in real ecosystems so far, we expect them to occur due to an accelerating growth and expansion of the human population. The land-use change associated with this increasing human population \citep{Kees2011,Ramankutty2016} will very likely lead to increasing rates of environmental change \citep{Walther2002,Joos2008} and subsequently to possible rate-induced critical transitions in ecosystems.\\
 In ecology, much effort has been devoted to the identification of alternative states to explain observed regime shifts related to changes in biodiversity or dominance of species in a given ecosystem by means of bifurcation and hysteresis behavior \citep{Rocha2015,Wernberg2016}. However, one can speculate that some of those observed regime shifts might in fact be rate-induced instead of bifurcation-induced transitions because the rate of the changing environment has rarely been taken into account. Consequently, the response of ecosystems to rates of environmental change has to be studied in more detail - in theory and the natural world - to better understand and possibly prevent unexpected transitions and regime shifts with dramatic ecological consequences.

\bibliographystyle{apalike} 
\bibliography{literature3}
\begin{appendix}
\appendix
\numberwithin{equation}{section}
\counterwithin{figure}{section}
\begin{appendices}
\section{The non-dimensionalized Rosenzweig-MacArthur predator-prey model}
\label{app_A}
The Rosenzweig-MacArthur predator-prey system with prey population density $x$ and predator population density $y$ is given by the following equations \citep{Rosenzweig1963}: 

\begin{align}
\label{eq1}
&\frac{dx}{d\tau} = ax\left(1-\frac{x}{K}\right) - \frac{\alpha x y}{1+\beta x}\\
\label{eq2}
&\frac{dy}{d\tau} = \gamma \frac{\alpha x y }{1+\beta x} - cy
\end{align}

employing a logistic growth of the prey with the maximum per-capita growth rate of the prey $a$ and the carrying capacity $K$ which determines the maximum population density of the prey in the equilibrium when the predator is absent. The functional response of the predator is modeled as a Holling-Type II function characterized by the maximum predation rate $\alpha/\beta$ and the half-saturation constant $1/ \beta$. The conversion efficiency $\gamma$ specifies the ratio between biomass increase and food uptake. The predator's  mortality is assumed to be proportional to its population density and, hence, represented by the linear term $-cy$.\\
The time $\tau$ is scaled by the per-capita growth rate of the prey population $a$ resulting in the
non-dimensional time $\hat{t} = a\tau$. Therefore, Eqs.~\eqref{eq1} and \eqref{eq2} become:

\begin{align}
&\frac{dx}{d\hat{t}} = x\left(1-\frac{x}{K}\right) - \frac{1}{a} \frac{\alpha x y}{1+\beta x}\\
&\frac{dy}{d\hat{t}} = \frac{\gamma}{a} \frac{\alpha x y }{1+\beta x} - \frac{c}{a} y.
\end{align}

The state variable of the prey $x$ is scaled as $x=\frac{u c}{\alpha \gamma}$ while the predator $y$ is scaled by $y = \frac{a}{\alpha}v$ leading to:

\begin{align}
&\frac{du}{d\hat{t}} = u\left(1-\frac{u c}{\alpha \gamma K}\right) -  \frac{ u v}{1+\beta \frac{u c}{\alpha \gamma}}\\
&\frac{dv}{d\hat{t}} = \frac{c}{a} \left( \frac{ uv }{1+\beta \frac{u c}{\alpha \gamma}} - v\right)
\end{align}

With the parameters $\eta = \frac{\beta c}{\alpha \gamma}$, $\phi = \frac{c}{\alpha \gamma K}$ and
$\kappa = \frac{c}{a}$, the system can be formulated in non-dimensional time $t$ as well as non-dimensional state variables of prey $u$ and predator $v$ as follows:

\begin{align}
\label{eq:final_app}
&\frac{du}{d\hat{t}} = u(1-\phi u) - \frac{uv}{1+\eta u}\\
\label{eq:final2_app}
&\frac{dv}{d\hat{t}} = \kappa  \left(\frac{uv}{1+\eta u}  - v\right).
\end{align}

The dimensionless parameter $\phi$ is proportional to the inverse of the carrying capacity, while $\eta$ denotes the dimensionless handling time of the predator. The parameter $\kappa$ describes the relation between the characteristic time scales of predator and prey. In general, the prey has a faster life cycle than the predator, and therefore, $0<\kappa \ll 1$. Hence, the non-dimensional time $\hat{t}$ would be called the fast time in fast-slow system theory as outlined in more detail in the following appendix 'Fast-slow system theory applying to the fast-slow Rosenzweig-MacArthur model'.
\newpage
\section{Fast-slow system theory applied to the fast-slow Rosenzweig-MacArthur model}
\label{app_B}
The fast-slow Rosenzweig-MacArthur predator-prey system consists of a fast evolving prey population $u$ and a slower reproducing predator population $v$ which evolve on different time scales whose ratio is expressed by the parameter $\kappa$.  Since it is assumed that the life times of predator and prey differ substantially $\kappa$ is given by a small number  $0<\kappa\ll 1$. In fast time $\hat{t}$, the system is given by:

\begin{align}
\label{eq:fast}
&\frac{du}{d\hat{t}} = u(1-\phi u) - \frac{uv}{1+\eta u}\\
\label{eq:fast2}
&\frac{dv}{d\hat{t}} = \kappa \left( \frac{u}{1+\eta u} - v \right)
\end{align}

and can be written in slow time $t = \hat{t} \kappa$:

\begin{align}
\label{eq:slow}
\kappa&\frac{du}{dt} = u(1-\phi u) - \frac{uv}{1+\eta u}\\
\label{eq:slow2}
&\frac{dv}{dt} =   \frac{uv}{1+\eta u} - v.
\end{align}

\begin{figure}[H]
\centering
\includegraphics[scale=0.6]{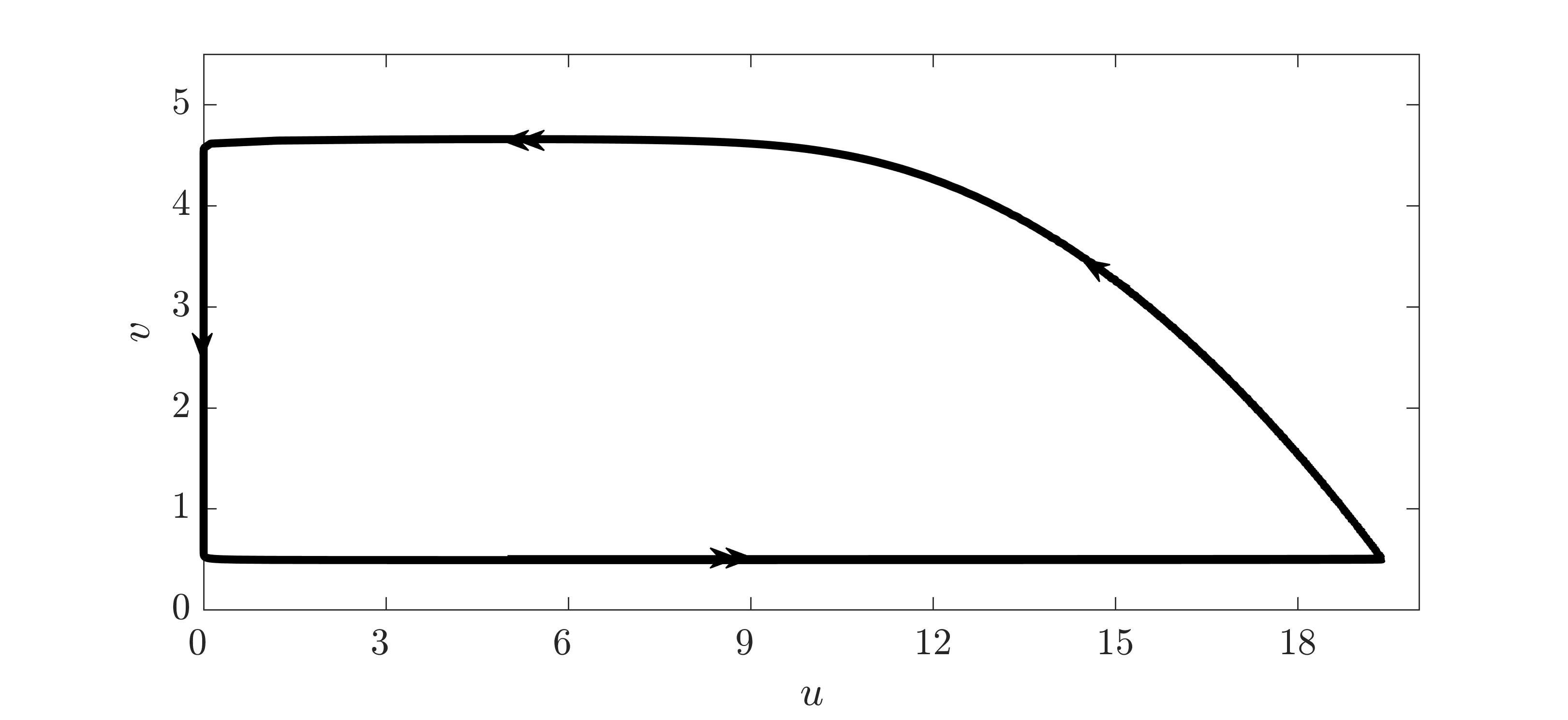}
\caption{Solution of the slow-fast predator-prey system (black trajectory), given by Eqs.~\eqref{eq:slow} and \eqref{eq:slow2}, whose dynamics can be characterized as a mixture of fast (two arrows) and slow (one arrow) motion.}
\label{fig:fast_slow_motion}
\end{figure}

For example, the prey population $u$ evolves ten times faster as the predator population $v$ when the parameter $\kappa=0.1$. Most of the time, the dynamics of fast-slow systems is determined by slow motion (one arrow) which is, at times, interrupted by short fast transitions (two arrows) as shown in figure \ref{fig:fast_slow_motion}. 

In fast-slow system theory, this mixture is analyzed by studying the fast and the slow motion in their limits to get an idea of the dynamics of the full system. \\

\begin{figure}[H]
\centering
\includegraphics[scale=0.6]{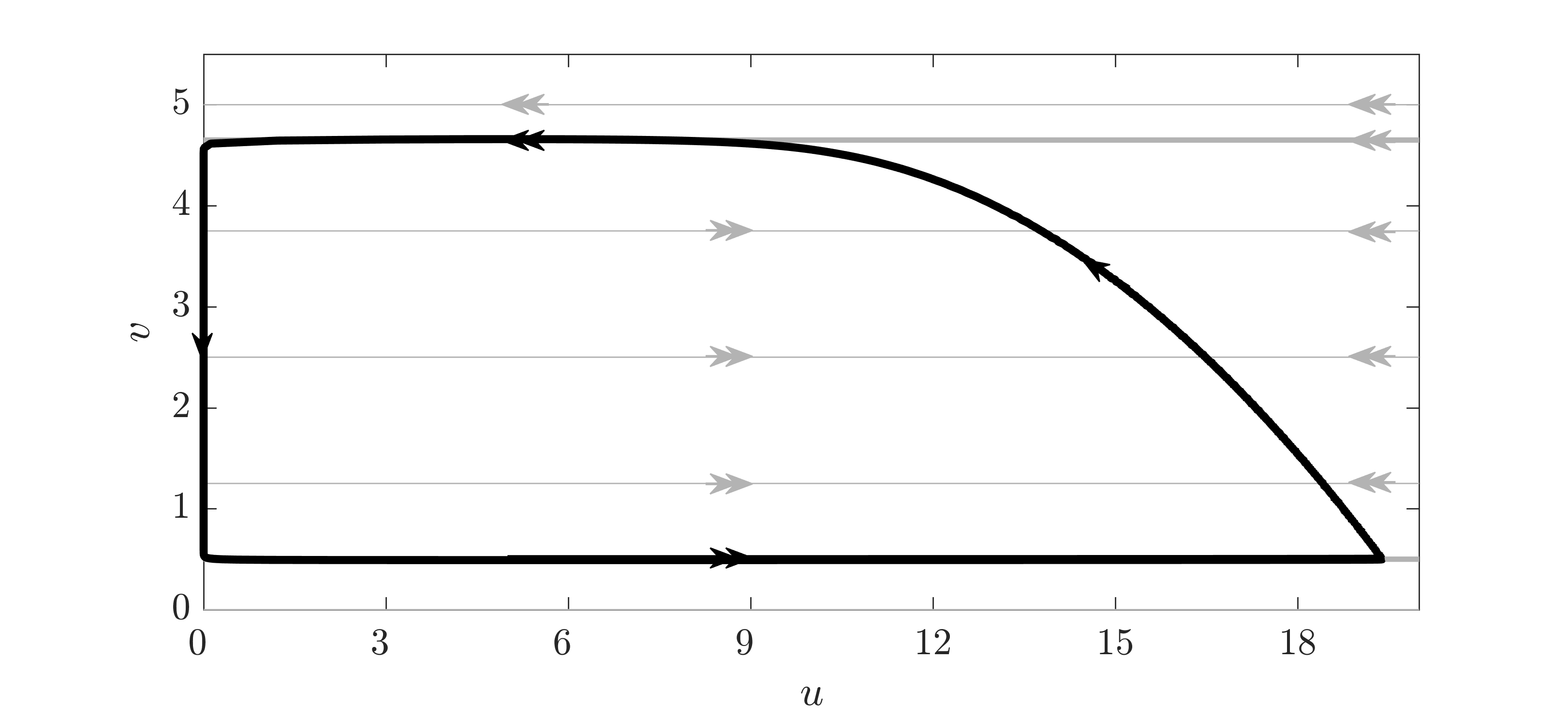}
\caption{The fast flow of the slow-fast predator-prey system (black trajectory, two arrows), given by Eqs.~\eqref{eq:slow} and \eqref{eq:slow2}, follows the solutions of the layer systems, given by Eqs. \eqref{eq:layer_system} and \eqref{eq:layer_system2} (horizontal gray lines).}
\label{fig:layer_system}
\end{figure}

When $\kappa \rightarrow 0$, the time-scaled system converges during fast segments to solutions of the fast subsystem called \textit{layer system} (see fig. \ref{fig:layer_system}):

\begin{align}
\label{eq:layer_system}
&\frac{du}{d\hat{t}} = u(1-\phi u) - \frac{uv}{1+\eta u}\\
\label{eq:layer_system2}
&\frac{dv}{d\hat{t}} = 0
\end{align}

which describes the evolution of the fast variable $u$ for fixed $v$ in fast time $\hat{t}=\frac{t}{\kappa}$.

Whereas during slow segments, trajectories of the slow-fast predator-prey system converge to solutions of the slow subsystem called \textit{reduced system}:

\begin{align}
\label{eq:reduced_system}
0 &= u(1-\phi u) - \frac{uv}{1+\eta u}\\
\label{eq:reduced_system2}
\frac{dv}{dt} &=   \frac{u}{1+\eta u} -v.
\end{align}

The reduced system describes the evolution of the slow variable $v$ in slow time $t$ on the so-called \textit{critical manifold} $S_0$ which is given by the set:

\begin{equation}
\label{critical_manifold}
S_0(\phi,\eta) = \left\lbrace(u,v) \in \mathbb{R}^2 : u\left((1-\phi u) - \frac{v}{1+\eta u}\right)=0, u \geq 0,v \geq 0\right\rbrace,
\end{equation}

which consists of two components, the straight component $u=0$ and the folded component $v = (1-\phi u)(1+\eta u)$. The folded component has a fold
tangent to the fast $u$-direction at the point
$$
F(\phi,\eta) = (u_F,v_F) = \left(\frac{\eta-\phi}{2\eta\phi },\frac{(\eta + \phi)^2}{4\eta\phi} \right), 
$$

\begin{figure}[H]
\centering
\includegraphics[scale=0.6]{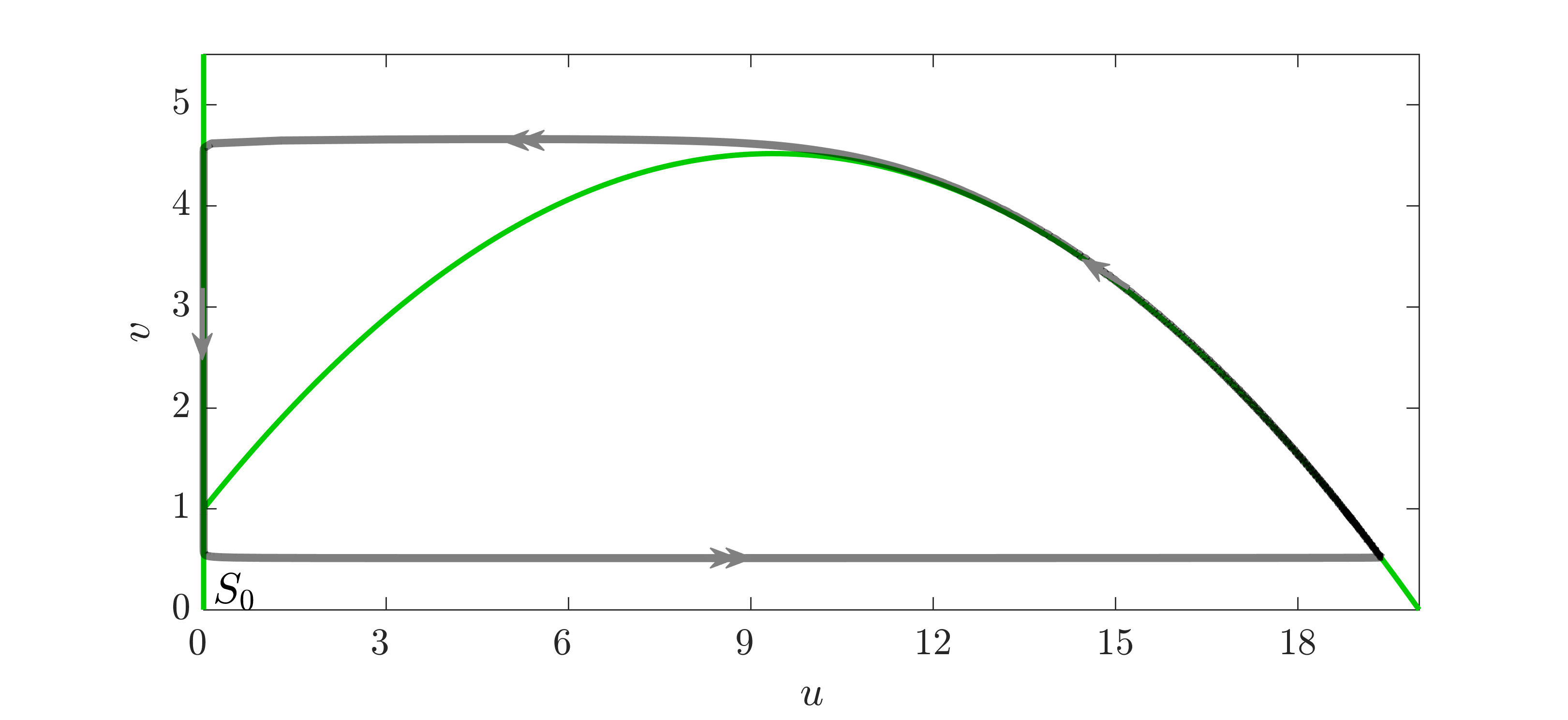}
\caption{The slow flow of the slow-fast predator-prey system (gray trajectory, one arrow), given by Eqs. \eqref{eq:slow} and \eqref{eq:slow2}, approaches the critical manifold $S_0$ (green), given by Eq. \eqref{critical_manifold}.}
\label{fig:reduced_system}
\end{figure}

The critical manifold $S_0$ (green) approximates the slow dynamics of the fast-slow system as shown in figure \ref{fig:reduced_system}.

\begin{figure}[H]
\centering
\includegraphics[scale=0.6]{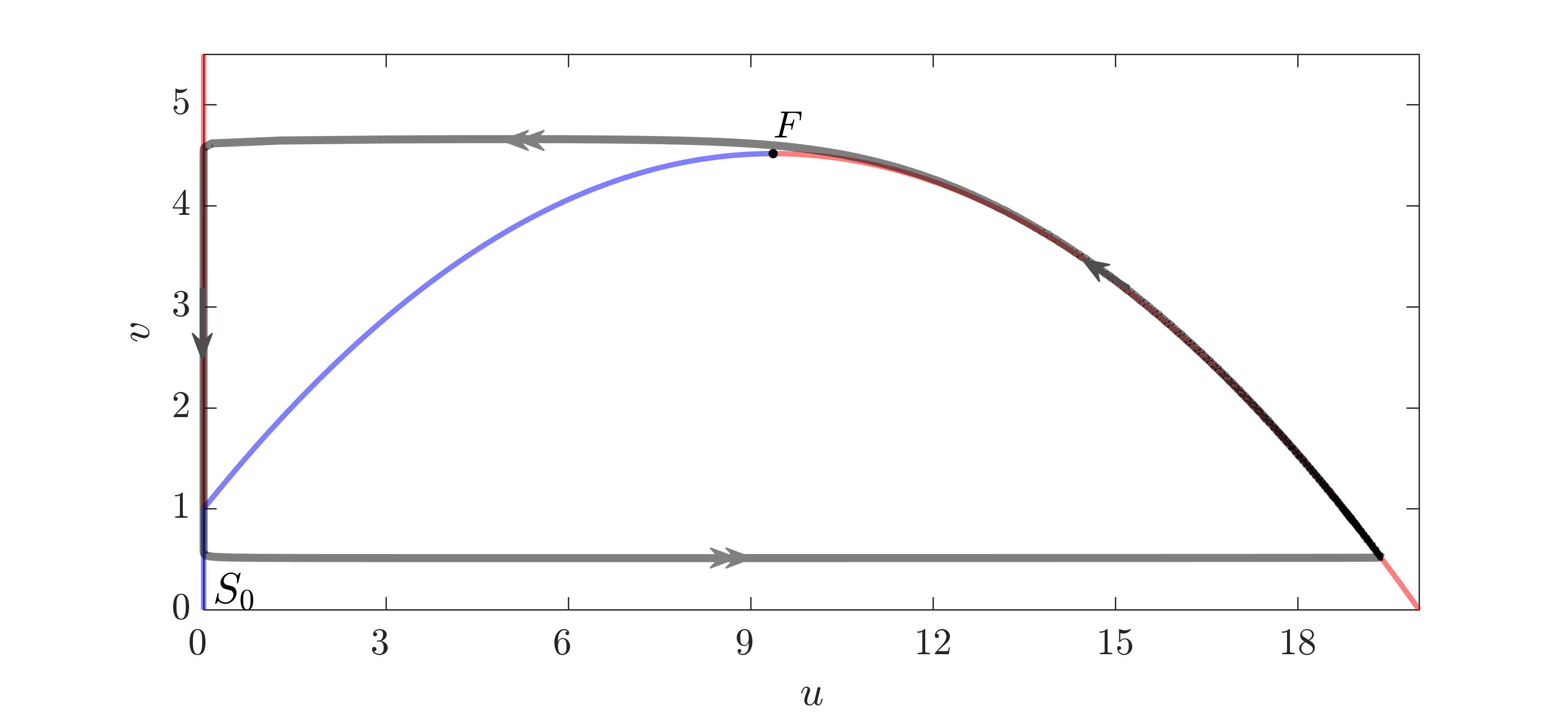}
\caption{The critical manifold $S_0$ can be divided into stable (red) and unstable parts (blue) and a fold (non-normally hyperbolic) point $F$ (black dot).}
\label{fig:stable_unstable_branches_fold}
\end{figure}

Additionally, the critical manifold $S_0$ is the set of equilibria of the layer system \eqref{eq:layer_system}--\eqref{eq:layer_system2} which are given by its intersection points with the critical manifold $S_0$. Consequently, the critical manifold $S_0$ defines an interface between the slow flow, given by the reduced system \eqref{eq:reduced_system}--\eqref{eq:reduced_system2}, and the fast flow, given by layer system \eqref{eq:layer_system}--\eqref{eq:layer_system2}. \\
Moreover, a subset of the critical manifold $S_h \subseteq S_0$ is called normally hyperbolic if all $(u,v) \in S_h$ are hyperbolic equilibria of the layer system which means that the eigenvalues of the Jacobian with respect to the fast variable $u$ have no zero real part. This is valid for all points on the critical manifold $S_0$ except for the fold $F$ which is a non-normally hyperbolic equilibrium of the layer system. The hyperbolic equilibria of the layer system \eqref{eq:layer_system}--\eqref{eq:layer_system2} are stable when $\lambda<0$ and unstable when $\lambda>0$. As shown in figure \ref{fig:stable_unstable_branches_fold}, stable parts of the critical manifold $S_a$ (red) attract trajectories in phase space whereas unstable parts $S_r$ (blue) repel them towards stable parts. 

\newpage
\section{Overconsumption by the predators causing the temporary collapse of the prey}
\label{app_temporary_collapse}
\begin{figure}[H]
\centering
\includegraphics[scale=0.7]{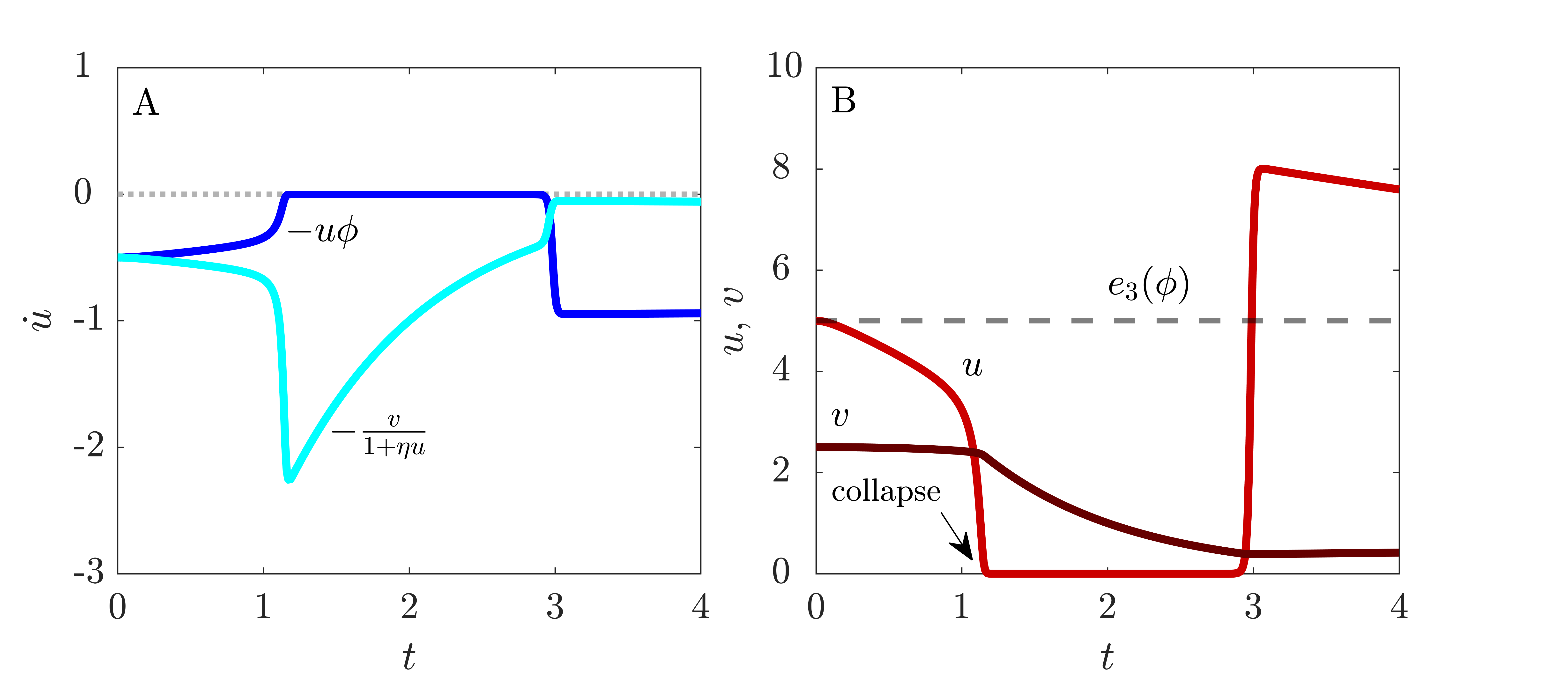}
\caption{(A): Decline of the prey population density \mbox{$u$} due to per-capita intraspecific competition \mbox{$-\phi u$} (dark blue) and per-capita predation pressure \mbox{$v/(1+\eta u)$} (light blue). (B): Time evolution of prey density $u$ (light red) and predator density \mbox{$v$} (dark red) of the ramped system \mbox{\eqref{rampedRM1}--\eqref{phi}}. The moving stable equilibrium \mbox{$e_3(\phi)$} is denoted as gray dashed line. At a particular time, the prey population is confronted to such a massive overconsumption by the predator population and undergoes a rate-induced critical transition leading to the collapse of the prey population. Parameters: \mbox{$\kappa = 0.01$, $\eta = 0.8$, $\phi_0 = 0.1$,}  \mbox{ $r=0.006$, $u_0 = (1-\eta)^{-1}$, $v_0 = (1-\phi_0 u_0)(1+\eta u_0)$. }}
\label{fig:bio_interp}
\end{figure}

To study the biological mechanisms which lead to the collapse of the prey population in figure {\ref{fig:ramping_process}}B, we focus on the two mechanisms which reduce the prey population density $u$ in the ramped system \mbox{\eqref{rampedRM1}--\eqref{phi}}: the intraspecific competition represented by the term \mbox{$-\phi u^2$}, and predation represented by the term \mbox{$-uv/(1+\eta u)$}.
In the following, we analyze whether it is the increasing intraspecific competition, a growing predation pressure, or a combination of both effects, that is responsible for the collapse of the prey population. Hence, we plot the per-capita loss of the prey population density due to both mechanisms in time separately in figure {\ref{fig:bio_interp}}A, together with the corresponding time evolution of the prey density \mbox{$u$} and predator density \mbox{$v$} in figure {\ref{fig:bio_interp}}B.

As seen in figure {\ref{fig:bio_interp}}A, both mechanisms -- intraspecific competition (blue) and predation (light blue) -- cause the prey population to decline. 
However, their contribution to the total per-capita loss is changing in time. As the resources are reduced in time, the prey population initially experiences a large per-capita intraspecific competition which then diminishes over time because lower population densities reduce the per-capita competition for resources. Concurrently, the predation pressure on the shrinking prey population intensifies because the population density of the predators declines very slowly. Due to their slower evolution in time, predators are not able to adapt quickly enough to the shrinking prey population. As a result, a small prey population is confronted with a large predator population, leading to an overwhelming predation pressure on the prey and to a temporary collapse of the prey population.

\newpage
\section{Dependence on the predator's handling time $\eta$}
\label{app_impact_eta}
\begin{figure}[H]
\centering
\includegraphics[scale=0.7]{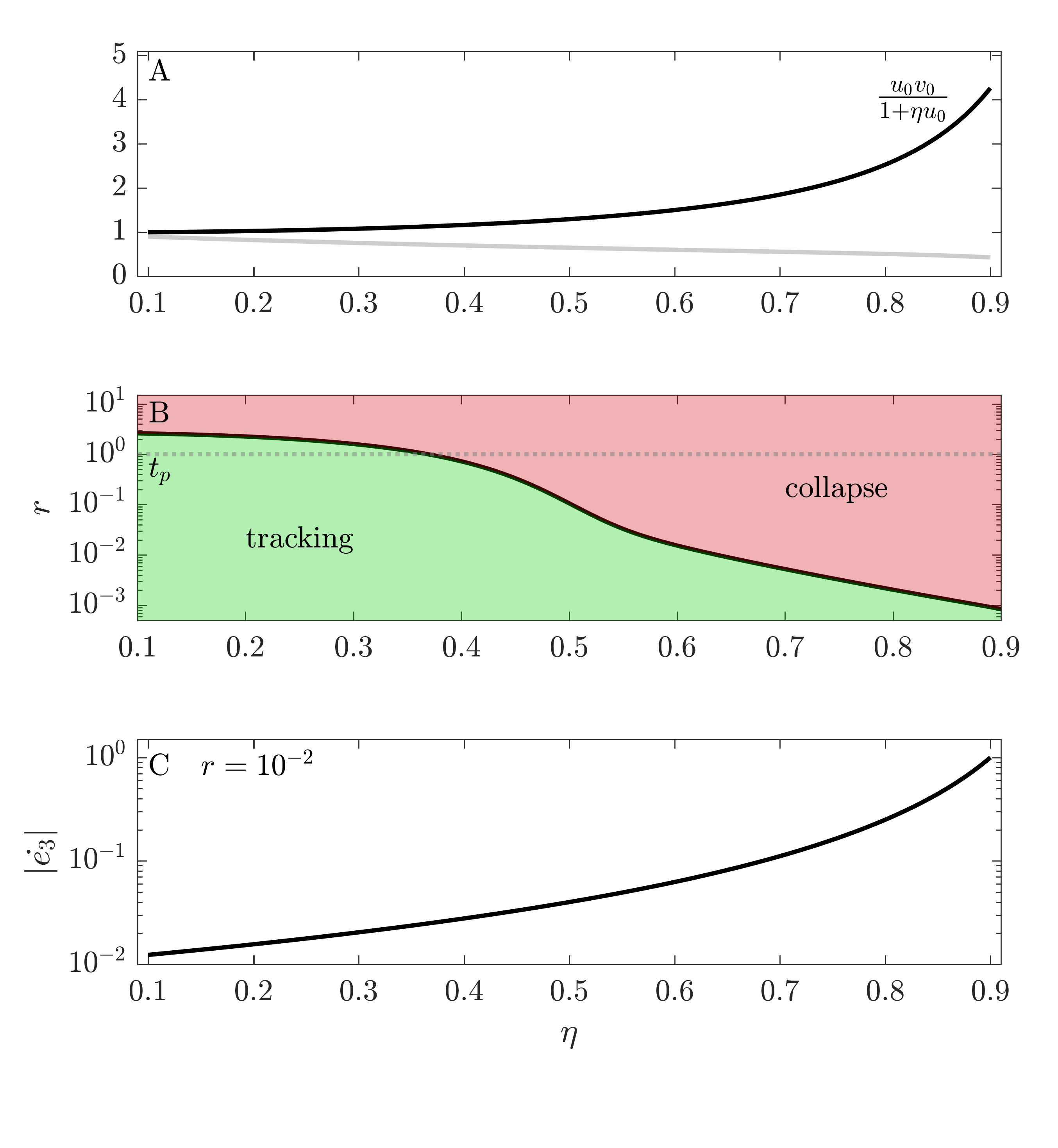}
\caption{(A): Initial predation pressure \mbox{$\frac{u_0v_0}{1+\eta u_0}$} (black) and initial per-capita predation pressure \mbox{$\frac{v_0}{1+\eta u_0}$} (gray) depending on the predator's handling time $\eta$. (B):  Parameter combinations \mbox{$(\eta,r)$} within the red/green region lead to collapse/tracking. The critical rate \mbox{$\hat{r}_{crit}$} and predator's timescale are denoted as black solid line and gray dotted line. (C): Speed of the moving stable equilibrium \mbox{$|\dot{e_3}|=\frac{r}{(1-\eta)^2}$} depending on the predator's handling \mbox{$\eta$} at the fixed rate \mbox{$r=10^{-2}$}. Parameters: \mbox{$\kappa = 0.01$, \mbox{$\epsilon=10^{-6}$},  $u_0=(1-\eta)^{-1}$, $\phi_0=\phi_{min,\epsilon}+0.005$, $v_0=(1-\phi_0 u_0)(1+\eta u_0)$}, \mbox{$\phi_{min\epsilon} = \frac{\eta(1-\eta)}{(1+\eta)+\epsilon}$, $\phi_{max,\epsilon}=1-\eta-\epsilon$.}}
\label{fig:affect_of_eta}
\end{figure}

In the previous sections, we have often mentioned that the initial predation pressure is the determining factor leading to a rate-induced critical transition of the ramped predator-prey system \mbox{\eqref{rampedRM1}--\eqref{phi}}. Hence, we study in the following how the initial predation pressure \mbox{$u_0v_0/(1+\eta u_0)$} determined by the initial prey and predator density \mbox{$u_0$}
and \mbox{$v_0$} respectively, as well as the predator's handling time \mbox{$\eta$} affects the occurrence of rate-induced critical transition in the ramped system. To this end, we compute the critical rate \mbox{$\hat{r}_{crit}$} at which the prey population collapses for different predator's handling time \mbox{$\eta$}. \\
The handling time of the predator \mbox{$\eta$} describes the time the predator needs to hunt, to kill and to digest its prey. 
So far, we have only considered a large predator population with inefficient individuals \mbox{$\eta=0.8$}. Figure {\ref{fig:affect_of_eta}}A demonstrates that such a predator population has a larger initial predation pressure on the whole prey population than a predator population characterized by a shorter handling time (black line). This seems counter-intuitive because a shorter handling time \mbox{$\eta$} increases the per-capita predation pressure on the prey population (gray line). However, a shorter handling time \mbox{$\eta$} means also that the initial predator population density \mbox{$v_0$} needs to be small to ensure a stable coexistence of predator and prey in the stable equilibrium \mbox{$e_3(\phi)$}.
In the following, we study how the occurrence of rate-induced critical transitions in the ramped system changes when the individuals of the predator population become more efficient - i.e. that their handling time \mbox{$\eta$} shortens.

From an ecological point of view one often considers a minimal prey population density below which the probability of extinction due to demographic or environmental noise becomes very high \mbox{\citep{Liephold2003}}. We are interested in the critical rate \mbox{$\hat{r}_{crit}$} at which the prey population density \mbox{$u$} suddenly drops below a minimal conservation population density $u_e$ during rate-induced tipping (see appendix {\ref{app_D}} for a more mathematically approach of the critical rate). We compute numerically this critical rate for different handling times \mbox{$\eta \in [0.1\; 0.9]$} and choose the minimal conservation population density $u_e = 0.2$, which amounts to approximately one fifth of the smallest initial prey density $u_0$ in the stable equilibrium \mbox{$e_3(\phi_0)$} for \mbox{$\eta=0.1$}. \\
In the beginning of the simulation, the ramped system is in equilibrium \mbox{$e_3(\phi_0)$} close to the fold \mbox{$F(\phi)$: $u_0 = (1-\eta)^{-1}$, $\phi_0 = \phi_{min}+0.005$} and \mbox{$v_0 = (1-\phi_0 u_0)(1+\eta u_0)$}. In this situation, the initial state of the ramped system is highly prone to rate-induced critical transitions (very high predation pressure). \\
Figure {\ref{fig:affect_of_eta}}B shows the parameter space of the rate of environmental change \mbox{$r$} and the predator's handling time $\eta$ in which the black solid line marks the critical rate \mbox{$\hat{r}_{crit}$} for the corresponding handling time of the predator \mbox{$\eta$}. The longer the handling time \mbox{$\eta$}, the smaller the critical rate \mbox{$\hat{r}_{crit}$}. This is due to the fact that a longer handling time \mbox{$\eta$} implies a faster moving stable equilibrium \mbox{$|\dot{e_3}| = \frac{r}{(1-\eta)^2}$} at a fixed rate \mbox{$r$} (see fig. {\ref{fig:affect_of_eta}}C). Consequently, a
nudge in
the fast direction $u$ causes the system to loose track of the moving stable equilibrium \mbox{$e_3(\phi)$}. In ecological terms, the system is exposed to a lower initial predation pressure when the predator's handling time shortens. For this reason, a higher intraspecific competition due to a higher rate of environmental change is needed to trigger a collapse of the prey population. \\
In figure {\ref{fig:affect_of_eta}}B, the black solid line divides the parameter space into the (green) \mbox{{\em tracking}} and (red) \mbox{{\em collapse}} region. The collapse region expands to smaller critical rates the more inefficient the predator gets. Since inefficient predators need larger predator population densities to track and maintain the equilibrium with the prey, they are more sensitive to perturbations. In this case, already very slow changes in the environment would lead to a loss of tracking and, hence to a rate-induced critical transition. Even environmental changes which are several orders of magnitude slower than the slowest time scale of the ecosystem, which is given by the 
rescaled predator lifetime \mbox{$t_p = 1$}, can result in rate-induced critical  transitions. Note further, that this relationship between handling time $\eta$ and critical rate \mbox{$\hat{r}_{crit}$} is highly nonlinear as depicted in figure~{\ref{fig:affect_of_eta}}B.

\newpage

\newpage

\section{The singular canard and the desingularized system}
\label{app_C}

To find the singular canard in the ramped Rosenzweig-MacArthur predator-prey system, given by Eqs. \eqref{app_rampedRM1}--\eqref{app_phi}, we have to study the fast flow $u$ in slow time $t$ on the critical manifold $S_0$. The ramped system is given by the following equations:

\begin{align}
\label{app_rampedRM1}
\kappa \frac{du}{dt} &= u(1-\phi u) - \frac{uv}{1+\eta u} \\
\label{app_rampedRM2}
\frac{dv}{dt} &= \frac{uv}{1+\eta u} - v \\
\label{app_phi}
\frac{d\phi}{dt} &= r
\end{align}

with time scale separation $\kappa$, handling time of the predator $\eta$ and $\phi \in [\phi_{min,\epsilon}\;\phi_{max,\epsilon}]$. The critical manifold $S_0$ of the ramped system is given by the set:

\begin{equation}
\label{critical_manifold_3d}
S_0 = \left\lbrace (u,v,\phi) \in \mathbb{R}^3 : u(1-\phi u) - \frac{uv}{1+\eta u}=0 \right\rbrace
\end{equation}
which can be written as a graph over $u$ and $\phi$ when $u\neq 0$
\begin{equation}
\label{graph_S0}
v=(1-\phi u)(1+\eta u)
\end{equation}

The folded component has a fold tangent to the fast u-direction at the point

\begin{equation}
F(\phi,\eta) = (u_F,v_F) = \left(\frac{\eta-\phi}{2\phi\eta},\frac{(\eta+\phi)^2}{4\eta\phi}\right).
\end{equation}

The flow on the critical manifold $S_0$ is determined by the so-called \textit{reduced system} when $\kappa \to 0$:

\begin{align}
\label{app_reduced_system}
0 &= u(1-\phi u) - \frac{uv}{1+\eta u} := f(u,v,\phi,\eta) \\
\label{app_reduced_system2}
\frac{dv}{dt} &= \frac{uv}{1+\eta u} - v \\
\label{app_reduced_system3}
\frac{d\phi}{dt} &= r
\end{align}

Differentiating the algebraic constraint \eqref{app_reduced_system} with respect to the slow time $t$ using the chain rule leads to an expression of the fast flow $u$ in slow time $t$ on the critical manifold $S_0$:

\begin{align}
\frac{du}{dt} &= -\frac{\frac{\partial f}{\partial v}\frac{dv}{dt}+\frac{\partial f}{\partial \phi}\frac{d\phi}{dt}}{\frac{\partial f}{\partial u}}\\
\label{app_fast_flow_in_slow_time}
\frac{du}{dt} &= \frac{\frac{u^2v}{(1+\eta u)^2}-\frac{uv}{1+\eta u} + u^2r}{1 - 2\phi u - \frac{v}{(1+\eta u)^2}}.
\end{align}

Replacing Eq. \eqref{app_reduced_system} by the formulation for the fast flow in slow time $\frac{du}{dt}$, given by Eq. \eqref{app_fast_flow_in_slow_time}, leads to:

\begin{align}
\label{app_reduced_fast_flow_in_slow_time}
\frac{du}{dt} &= \frac{\frac{u^2v}{(1+\eta u)^2}-\frac{uv}{1+\eta u} + u^2r}{1 - 2\phi u - \frac{v}{(1+\eta u)^2}}\\
\label{reduced_system2fs}
\frac{dv}{dt} &= \frac{uv}{1+\eta u} - v \\
\label{reduced_system3fs}
\frac{d\phi}{dt} &= r.
\end{align}

Further, we can project the dynamics of the reduced system, given by Eqs. \eqref{app_reduced_fast_flow_in_slow_time}--\eqref{reduced_system3fs}, onto the two-dimensional critical manifold $S_0$ by using Eq.~\eqref{graph_S0}. This results in the so-called two-dimensional \textit{projected reduced system}:

\begin{align}
\label{app_projected_reduced_system}
\frac{du}{dt} &= \frac{u(1-\phi u)-(1-\phi u)(1+\eta u) + u(1+\eta u)r}{2\phi \eta(u_F-u)}:=\frac{\Lambda(u,\phi,\eta,r)}{2\phi\eta(u_F-u)}\\
\label{app_projected_reduced_system2}
\frac{d\phi}{dt} &= r.
\end{align}

At the fold $F$ of the critical manifold $S_0$, the term $2\phi\eta(u_F-u) = 0$ because $u=u_F$ which results in a blow up of the fast flow in slow time $\frac{du}{dt}$ at the fold $F$. Therefore, solutions of the projected reduced system are not able to cross the fold $F$ from the stable parts towards the unstable parts of the critical manifold $S_0$ away from the folded singularity $FS$ (see fig. \ref{fig:folded_canard}A). But when simultaneously the term in the nominator $\Lambda=0$ vanishes, a \textit{singular canard trajectory} can cross the fold $F$ with finite speed via the folded singularity equilibrium $FS$. In the ramped Rosenzweig-MacArthur predator-prey system, given by Eqs. \eqref{app_rampedRM1}--\eqref{app_phi}, the singular canard is the threshold separating initial conditions on the stable part of the critical manifold that undergo a rate-induced critical transition from those that track the moving stable equilibrium $e_3(\phi)$ (see fig. \ref{fig:dependence_on_ini_with_canard}). \\
To find the singular canard, we have to determine the folded singularity $FS$. As a consequence, we have to analyze the dynamics of the projected reduced system close to the fold $F$. As mentioned above, the fast flow on the critical manifold $\frac{du}{dt}$ is not defined at the fold $F$ due to the division by zero. For this reason, a mathematical trick is used, called \textit{desingularization}, which removes the term in the denominator $2\phi\eta(u_F-u)$ and reverses the time $t$ on the unstable part of the critical manifold $S_0$ by rescaling the time $t = -2\phi\eta(u_F-u)s$.\\
The \textit{desingularized system} can be written as follows:

\begin{align}
\label{des}
\frac{du}{ds} &= u(1-\phi u)-(1-\phi u)(1+\eta u) + u(1+\eta u)r = \Lambda(u,\phi,\eta,r)\\
\label{des2}
\frac{d\phi}{ds} &= 2\phi\eta(u_F-u)r.
\end{align}

Since the folded singularity $FS$ is determined by $\Lambda=0$ and $2\phi\eta(u_F-u)r=0$, $FS$ corresponds to an equilibrium of the desingularized system. Therefore, we have to compute the equilibria of the desingularized system to find the folded singularity $FS$:

\begin{align}
\label{des_eq_0}
0 &= \frac{u^2(1-\phi u)}{1+\eta u}-(1-\phi u) +u^2r\\
\label{des2_eq_0}
0 &= 2\phi\eta(u_F-u)r.
\end{align}

The only equilibrium of the desingularized system is the folded singularity $FS$ given by the coordinates $u_{FS}$ and $\phi_{FS}$:

\begin{align}
u_{FS} &= \frac{\eta + \sqrt{(1 + r - \eta)^2 + 8\eta r}-r-1}{4\eta r}\\
\phi_{FS} &= \frac{\eta}{2\eta u_{FS} + 1}.
\end{align}

\begin{figure}[H]
\centering
\includegraphics[scale=0.55]{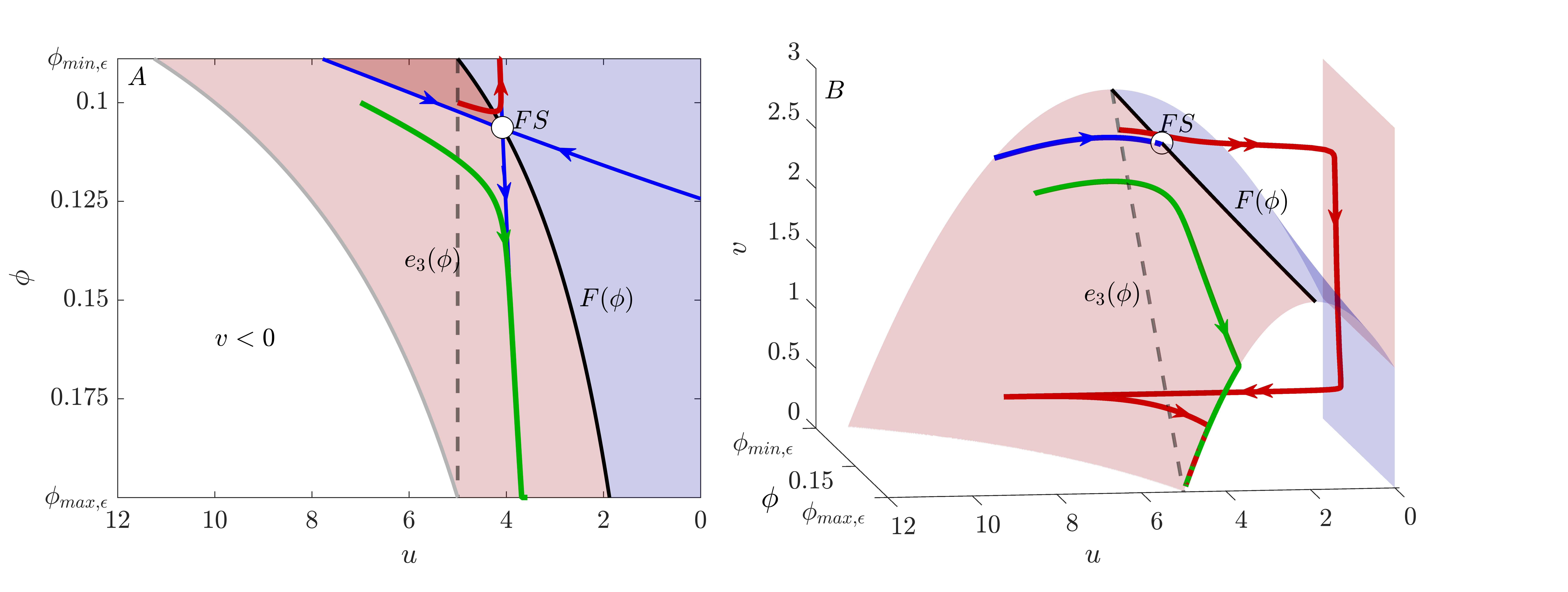}
\caption{(A) Phase portrait of the desingularized system, given by Eqs. \eqref{des} and \eqref{des2} with stable (red) and unstable part (blue) and the fold $F(\phi)$ (black solid line) of the critical manifold $S_0$. The pathway of the moving equilibrium $e_3(\phi)$ is denoted as gray dashed line. The stable manifolds of the folded saddle singularity ($FS$) (blue lines) are equivalent to the singular canard trajectory -  the tipping threshold. (Red) Initial conditions above the tipping threshold on the stable part of the critical manifold reach the fold and undergo a rate-induced critical transition wheres (green) initial conditions located below the tipping threshold track the moving stable equilibrium $e_3(\phi)$ (B): The singular canard (blue line) is the tipping threshold separating initial condition (red trajectory) that exhibit rate-induced tipping from tracking initial condition (green trajectory). Parameters: $\eta = 0.8$, $\phi_0 = 0.1$, $r = 0.006$, $u_0 = (1-\eta)^{-1}$ (red), $u_0 = 7$ (green), $v_0 = (1-\phi_0 u_0)(1+\eta u_0)$. }
\label{fig:phase_portrait_DES}
\end{figure}

The two eigenvalues of the folded singularity $FS$ are $\lambda_u<0$ and $\lambda_{\phi}>0$. Therefore, the folded singularity is a \textit{folded saddle} equilibrium of the desingularized system. According to \cite{Wieczorek2011}, the singular canard (the tipping threshold) is given by the stable invariant manifolds of the folded saddle equilibrium. \\
Figure \ref{fig:phase_portrait_DES}A shows the phase portrait of the desingularized system. The folded canard is given by the stable invariant manifolds of the folded saddle equilibrium $FS$ (horizontal blue lines). Remember, the time $s$ is reversed on the unstable part (blue) of the critical manifold $S_0$ in the desingularized system. To think in the 'real' time $t$, you have to reverse the arrows on the unstable part. The red trajectory crosses the fold $F(\phi)$, exhibits rate-induced tipping and proceeds towards $\phi_{min}$. Whereas the green trajectory proceeds towards $\phi_{max}$ by tracking the pathway of the moving equilibrium $e_3(\phi)$ (gray dashed line). In figure \ref{fig:phase_portrait_DES}B, the red trajectory starting 'above' the singular canard in the collapse-prone region undergoes a rate-induced critical transition. By contrast, the green trajectory starting below the singular canard tracks the moving stable equilibrium $e_3(\phi)$.

\newpage
\section{Two critical rates of environmental change}
\label{app_D}

\begin{figure}[H]
\centering
\includegraphics[scale=0.6]{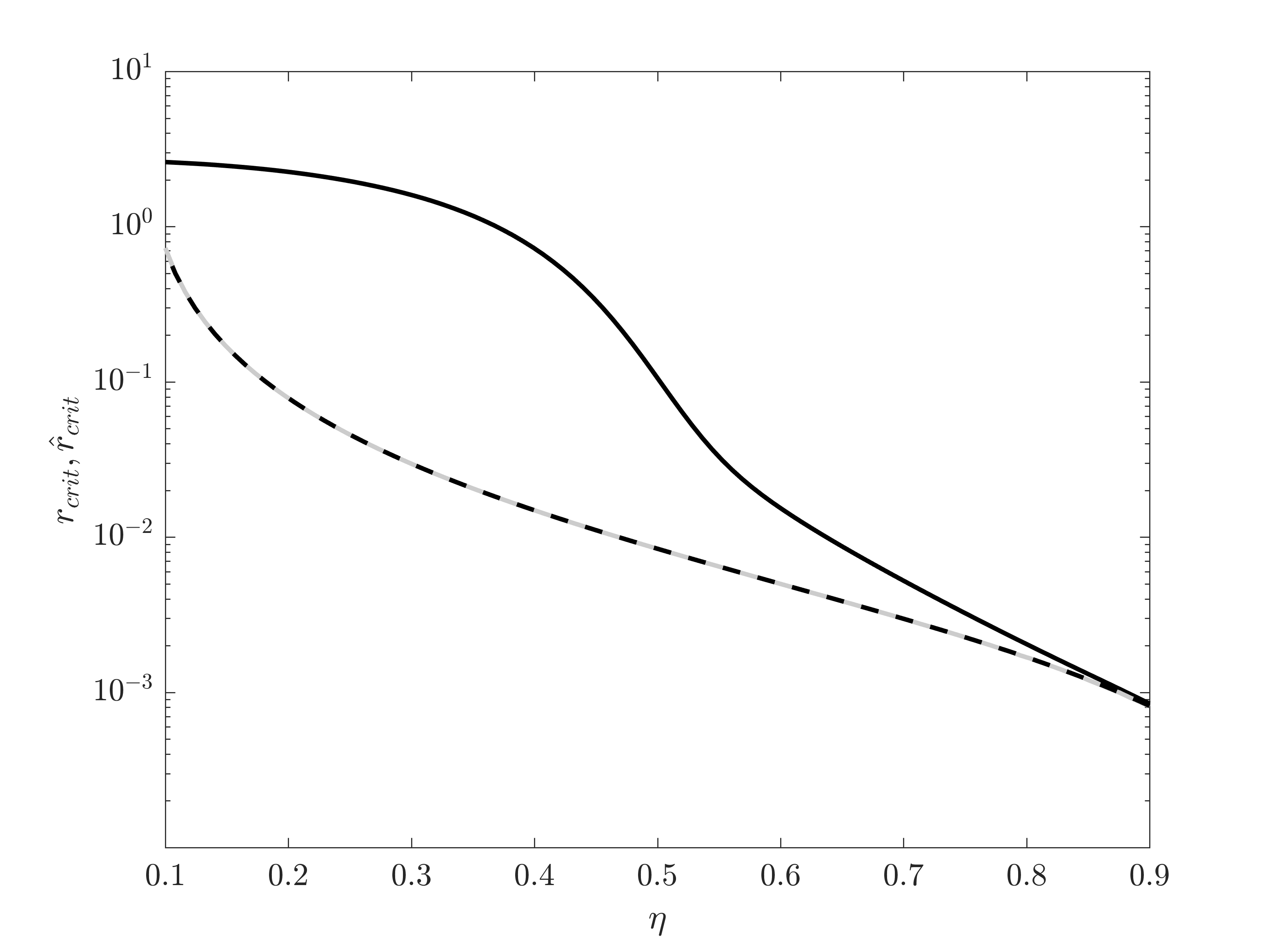}
\caption{The critical rates $r_{crit}$ and $\hat{r}_{crit}$ depending on the predator's handling time $\eta$. The black solid line denotes the critical rate $\hat{r}_{crit}$ as computed in section 'Dependence on the predator's handling time $\eta$' where the prey population collapses when $r=\hat{r}_{crit}$. The gray solid line marks the critical rate $r_{crit}$ approximated by Eq. \eqref{rcrit_eig} which means a rate-induced critical transition when $r=r_{crit}$ but not necessarily a collapse of the prey population. The back dashed line represents the critical rate $r_{crit}$ computed numerically to verify the approximation of the critical rate as given by Eq. \eqref{rcrit_eig}. Parameters: $u_0 = (1-\eta)^{-1}$, $\phi_0 = \phi_{min}+0.005$, $\eta \in [0.1\;0.9]$, $u_e = 0.2$.}
\label{fig:approx_rc}
\end{figure}

To approximate the critical rate of environmental change we can adapt two different approaches: a mathematical one and an ecological one. The mathematical approach of the critical rate $r_{crit}$ according to \cite{Wieczorek2011} is related to the computation of a tipping threshold (singular canard) by using the desingularized system \eqref{des}--\eqref{des2}. The tipping threshold separates initial states of the system into those that exhibit rate-induced tipping and those that track the moving equilibrium at the critical rate $r_{crit}$. The ecological approach of the critical rate $\hat{r}_{crit}$ includes an additional ecological condition which is useful when studying the sensitivity of populations to rate-induced tipping: the population density must fall below a critical conservation density $u_e$ during rate-induced tipping. We have already introduced the ecological approach of the critical rate $\hat{r_{crit}}$ in section 'Dependence on the predator's handling time $\eta$'. The critical rate $\hat{r_{crit}}$ (black line) in figure \ref{fig:approx_rc} is equivalent to the critical rate $\hat{r_{crit}}$ (black line) shown in figure \ref{fig:affect_of_eta}B. \\

According to \cite{Wieczorek2011}, the critical rate $r_{crit}$ at which a given initial condition ($u_0$,$\phi_0$) of the desingularized system undergoes a rate-induced critical transition can be approximated by using the eigenvector $w(r) = (w_1(r),w_2(r))^T$ corresponding to the negative eigenvalue $\lambda_1$ of the folded saddle equilibrium $FS(r) = (u_{FS}(r),\phi_{FS}(r))$. The stable eigenvector approximates the tipping threshold: the singular canard trajectory, close to the folded saddle equilibrium. Therefore, the critical rate $r_{crit}$ is the rate at which the stable eigenvector of the folded saddle equilibrium points exactly to the initial condition ($u_0$,$\phi_0$) of the desingularized system. When the rate $r$ increases such that $r>r_{crit}$, the location of the stable eigenvector changes in the way that the initial condition is located on the side of the singular canard where it exhibits rate-induced tipping. Notice, this approximation of the critical rate is only valid for initial conditions ($u_0$,$\phi_0$) of the desingularized system which are located close to the fold of the critical manifold due to the linearization.

\begin{equation}
\label{rcrit_eig}
\phi_0 - \phi_{FS}(r_{crit}) = \frac{w_2(r_{crit})}{w_1(r_{crit})}[u_0 - u_{FS}(r_{crit})]
\end{equation}

with:

\begin{align}
&u_{FS} = \frac{\eta + \sqrt{(1+r_{crit}-\eta)^2+8\eta r_{crit}} -r_{crit} -1}{4\eta r_{crit}}\\
&\phi_{FS} = \frac{\eta}{2\eta u_{FS} + 1}.
\end{align}

In the following, we compute the critical rate $r_{crit}$ according to Eq. \eqref{rcrit_eig} for different handling times $\eta \in [0.1\;0.9]$ of the predator. The initial conditions are given by: $u_0 = (1-\eta)^{-1}$ and $\phi_0 = \phi_{min}+0.005$. This means that the ramped system, given by Eqs. \eqref{app_rampedRM1}--\eqref{app_phi}, is in equilibrium $e_3(\phi_0)$ close to the fold $F(\phi)$ of the critical manifold $S_0$ and the lower boundary of the ramping interval $\phi_{min}$. Additionally, we check the approximation \eqref{rcrit_eig} by simulating the desingularized system, given by Eqs. \eqref{des} and \eqref{des2}, for the same initial conditions ($u_0$,$\phi_0$) and determine numerically the critical rate $r_{crit}$ at which the solution of the desingularized system reaches the fold of the critical manifold and undergoes a rate-induced critical transition. \\


Figure \ref{fig:approx_rc} shows the three resulting curves of the critical rate $r_{crit}$ depending on the predator's handling time $\eta$. The initial condition ($u_0$,$\phi_0)$ is identical in all three cases. The gray solid line represents the critical rate $r_{crit}$ according to the linearization, as given in Eq. \eqref{rcrit_eig}. The black dashed line shows the critical rate $r_{crit}$ computed numerically by simulating the desingularized system. Finally, the black curve denotes the critical environmental rate of change $\hat{r}_{crit}$ which implies a collapse of the prey population when $r=\hat{r}_{crit}$.\\

The critical rate $r_{crit}$ according to Eq. \eqref{rcrit_eig} (gray solid line) and the critical rate computed numerically (black dashed line) are almost identically. This clearly demonstrate that the critical rate $r_{crit}$ according to \cite{Wieczorek2011} is an excellent approximation of the critical rate when initial condition ($u_0$,$\phi_0$) is located close to the fold of the critical manifold. \\

The critical rates $\hat{r}_{crit}$ shown by the black solid line are higher compared to the critical rates $r_{crit}$ represented by the gray solid line particularly when the handling time of the predator $\eta$ is short. In this case, the moving equilibrium $e_3(\phi)$ moves slowly and (see sec. Population collapse due to a rate-induced critical transition), hence, the fast responding predator has a better change to 'catch up' with the moving equilibrium. As a consequence,  the faster response of the predator has to be balanced by a faster rate of the changing environment to finally end up in a population collapse. 


\end{appendices}
\end{appendix}
\end{document}